\documentclass[11pt]{article}
\usepackage{jheppub}

\usepackage{graphics}
\usepackage{amssymb}
\usepackage{amsfonts}
\usepackage{amsbsy}
\usepackage{amsmath}
\usepackage[vcentermath]{youngtab}
\usepackage{graphicx}
\usepackage{amscd}
\usepackage{epstopdf}
\usepackage{amsthm}
\usepackage{enumerate}

\newcommand{\cB}{\mathcal{B}}

\newcommand{\cF}{\mathcal{F}}
\newcommand{\cG}{\mathcal{G}}
\newcommand{\cH}{\mathcal{H}}

\newcommand{\cL}{\mathcal{L}}

\newcommand{\cO}{\mathcal{O}}

\newcommand{\cS}{\mathcal{S}}

\newcommand{\bC}{\mathbb{C}}

\newcommand{\bP}{\mathbb{P}}
\newcommand{\bQ}{\mathbb{Q}}
\newcommand{\bR}{\mathbb{R}}

\newcommand{\bZ}{\mathbb{Z}}

\DeclareMathOperator{\tr}{tr}

\def\ov{\over}
\def\lam{{\lambda}}

\def\eq#1{(\ref{#1})}

\def\Om{{\Omega}}

\def \lam {\lambda}
\def \om {\omega}
\def \ra {\rightarrow}

\def\s{{\sigma}}

\def\ts{{\widetilde \sigma}}

\def\tJ{{\widetilde J}}
\def\tB{{\widetilde B}}

\def\tK{{\widetilde K}}
\def\tg{{\widetilde g}}
\def\cn{{\mathcal{N}}}

\def\thy#1#2#3{\mathcal{T}_{#3} (#1, #1 + #3 #2)}
\def\thyN{\mathcal{T}_3(N,N)}
\def\thyNkkp{\mathcal{T}_3 (N+3k,N+3k')}

\newcommand{\be}{\begin{equation}}
\newcommand{\ee}{\end{equation}}
\newcommand{\bea}{\begin{equation} \begin{aligned}}
\newcommand{\eea}{\end{aligned} \end{equation}}

\newcommand{\bln}{\begin{align}}
\newcommand{\eln}{\end{align}}
\newcommand{\bst}{\begin{split}}
\newcommand{\est}{\end{split}}
\newcommand{\bi}{\begin{itemize}}
\newcommand{\ei}{\end{itemize}}
\newcommand{\ben}{\begin{enumerate}}
\newcommand{\een}{\end{enumerate}}

\title{6D SCFTs and Gravity}
\author[a]{Michele Del Zotto,}
\author[b]{Jonathan J. Heckman,}
\author[c]{David R. Morrison,}
\author[d]{and Daniel S. Park}
\affiliation[a]{Jefferson Physical Laboratory\\
Harvard University\\
Cambridge, MA 02138, USA}
\affiliation[b]{Department of Physics\\
University of North Carolina\\
Chapel Hill, NC 27599, USA}
\affiliation[c]{Departments of Mathematics and Physics\\
University of California, Santa Barbara\\
Santa Barbara, CA 93106, USA}
\affiliation[d]{Simons Center for Geometry and Physics\\
Stony Brook University, \\
Stony Brook, NY 11794-3636, USA}

\emailAdd{delzotto {\rm at} physics.harvard.edu}
\emailAdd{jheckman {\rm at} email.unc.edu}
\emailAdd{drm {\rm at} math.ucsb.edu}
\emailAdd{dpark {\rm at} scgp.stonybrook.edu}

\abstract{ We study how to couple a 6D superconformal field theory (SCFT) to gravity. In F-theory, the models in question are obtained working on the supersymmetric
background $\mathbb{R}^{5,1} \times B$ where $B$ is the base of a compact
elliptically fibered Calabi-Yau threefold in
which two-cycles have contracted to zero size. When the base has orbifold singularities, we find that
the anomaly polynomial of the 6D SCFTs can be understood purely in terms of the intersection theory of fractional divisors: the anomaly coefficient vectors are identified with elements of the orbifold homology. This also explains why in certain cases, the SCFT can appear to contribute a ``fraction of a hypermultiplet'' to the anomaly polynomial.
Quantization of the lattice of string charges also predicts the existence of additional light states beyond
those captured by such fractional divisors. This amounts to a refinement to
the lattice of divisors in the resolved geometry. We illustrate these general
considerations with explicit examples, focusing on the case of F-theory on an elliptic Calabi-Yau threefold
with base $\mathbb{P}^2 / \mathbb{Z}_{3}$.
}

\begin{document}

\setcounter{tocdepth}{2}

\maketitle

%%%%%%%%%%%%%%%%%%%%%%%%%%%%%%%%%%%%%%%%%%%%
%
%              Body of paper
%
%%%%%%%%%%%%%%%%%%%%%%%%%%%%%%%%%%%%%%%%%%%%

%\newpage

\section{Introduction and summary} \label{s:introduction}

F-theory  \cite{Vafa,MV1,MV2} provides the broadest known arena for
constructing string vacua. Part of the utility of this formulation is that
many stringy ingredients such as seven-branes are automatically packaged in terms of elliptic fibrations and
Calabi-Yau geometry. For six-dimensional low energy effective field theories, this approach is particularly
powerful, and has led to a characterization of virtually all known string vacua.\footnote{For a recent review of the close correspondence between bottom-up and
top-down constraints on 6D theories, see \cite{TaylorTASI}.}

Recently, there has also been renewed interest in using F-theory as a tool to
systematically construct and study 6D superconformal field theories (SCFTs). Building
on earlier work (see e.g.\ \cite{Witten:1995ex, Witten:1995zh, WittenSmall,
Strominger:1995ac, Ganor:1996mu, SW6, Bershadsky:1996nu,
Blum:1997fw, instK3, Intriligator:1997dh, Brunner:1997gf, Hanany:1997gh}), there is now a complete classification of 6D SCFTs without
a Higgs branch \cite{HMV}, with steady progress on the classification of theories with a Higgs branch \cite{DHTV, Heckman, sixTOTAL}. In this paper we  consider the problem of coupling these systems to gravity.

An important assumption in much of the literature on 6D supergravity theories is that the matter fields organize according to
``conventional'' supermultiplets. This includes the gravity multiplet, the tensor multiplet and vector multiplet, as well as
(half) hypermultiplets. There are rather tight consistency conditions for the possible ways such ingredients can be combined. These requirements include
6D anomaly cancellation, as well the requirement that the lattice of BPS strings is properly quantized and unimodular\footnote{Recall that a unimodular lattice is a lattice equipped with an integer valued quadratic form such that its determinant is $\pm 1$}---see e.g. \cite{KMT2,SeibergTaylor}.

The situation with 6D superconformal subsectors is much less understood. First of all, it is quite clear that a consistent theory of gravity cannot be obtained by coupling to arbitrary SCFTs,\footnote{For example, it is well known that a sufficiently large number of M5-branes cannot be consistently recoupled to 6D gravity. Nevertheless, these are consistent field theories, and provide an M-theory realization of the
$A_N$ series of $(2,0)$ theories.} and here we begin the task of determining which of these models might still be consistent upon coupling
to gravity.

From the perspective of the 6D gravity theory, such a SCFT corresponds to the presence of a strongly coupled sector which exhibits approximate scale invariance (which is broken by Planck scale effects). We explain the way in which some of these models can indeed satisfy the anomaly cancellation and unimodularity constraints, giving rise to physically sound supergravity theories.  To accomplish this,
we shall use the F-theory description of 6D supergravity backgrounds. This consists of specifying a compact complex surface $B$, and the data of an elliptically
fibered Calabi-Yau threefold
%$T^2 \rightarrow X \rightarrow B$.
$X \rightarrow B$.
To generate the strongly coupled conformal subsectors of our 6D supergravity theory, we allow two-cycles in the base to degenerate to zero size. The strings obtained by wrapping D3-branes over such vanishing cycles then become tensionless, leading to the desired SCFTs.

It would be quite desirable to understand F-theory away from the limit in which the volume of all cycles in the base are
large. Indeed, a potential weaknesses in the ``large volume perspective'' is the absence of a systematic
$\alpha^{\prime}$ expansion. The effects of short distance physics can
often be recovered by taking various singular degeneration limits. In the physical theory, this corresponds to adding light degrees of freedom in the low energy effective field theory, namely D3-branes wrapping vanishing cycles. One might therefore ask whether one can provide an intrinsic formulation of an F-theory compactification away from
the large volume limit. The benefits of having such a formulation would be considerable. For one, it would allow one to dispense with the
assumption that there is a moduli space of vacua connecting different regimes of parameter space.\footnote{For example, in 4D $\mathcal{N} = 1$ vacua, there can be obstructions to motion on the geometric moduli space of a compactification,
trapping the theory at small volumes \cite{Cordova}. See also \cite{FUZZ} for a discussion of how the
open string metric may nevertheless remain at large volume, and the corresponding formulation in terms of
non-commutative geometry.} Another aim of this paper will be to take some preliminary steps in this direction. We
focus on the case of 6D F-theory vacua with eight real supercharges.
This is a particularly tractable example to study, because there are a number of universal strong consistency conditions. Additionally, there is typically a moduli space of vacua which will enable us to check our formalism by moving
back to the large radius limit. In more detail, we will be interested in giving a direct geometric formulation of 6D F-theory vacua where the (compact) base $B$ of the elliptic fibration
%$T^2 \rightarrow X \rightarrow B$
$X \rightarrow B$
contains orbifold
singularities of the type considered in \cite{HMV, DHTV}.

Now, in the absence of any superconformal subsystem, the 6D effective theory supports a set of strings which can couple to various two-form potentials. Geometrically, the resulting lattice of charges $\Lambda_{string}$ is identified with the second homology lattice of the F-theory base:\footnote{Throughout this work, we do not consider possible discrete torsional contributions to the lattice of charges.}
\begin{equation}
\Lambda_{string} \equiv H_2(B_{smth}, \mathbb{Z}).
\end{equation}
Moreover, the intersection theory of $H_2(B_{smth}, \mathbb{Z})$ completely captures the anomaly polynomials of conventional matter. In the case where $B$ has orbifold singularities corresponding to strongly coupled SCFTs, the second orbifold homology contains fractional divisors, that is divisors which have an intersection pairing
valued over the rational numbers, rather than the integers. Quite surprisingly, the intersection theory of the lattice
\begin{equation}
\Lambda_{frac} \equiv H_2(B_{orb}, \mathbb{Z})
\end{equation}
still matches the anomaly coefficients of the corresponding SCFTs! Thus superconformal theories really behave as ordinary matter for this aspect of its F-theory realization. Na\"{\i}vely, it is quite tempting, in view of this result, to expect that this correspondence persists, namely that the string charges fill out the corresponding homology lattice $H_2(B_{orb}, \mathbb{Z})$. As this contains fractional divisors, we might therefore expect the theory to
contain fractional strings from D3-branes wrapped over such divisors. Indeed, the calculation of the
anomaly polynomial for these theories leads to contributions which fill out fractions of those of hypermultiplets. This may appear puzzling: How can such fractional states be compatible with the condition of charge quantization? Much in the spirit of \cite{BanksSeiberg}, the
answer is that charge quantization \textit{predicts} the existence of \textit{additional} strings. Once we supplement the theory with these additional states, the full theory turns out to
obey charge quantization (as it must). For this to work, there must exist a refinement $\Lambda_{ref}$ of our lattice $\Lambda_{frac}$, and a surjective map $\Lambda_{ref} \rightarrow \Lambda_{frac}$: Only the states spanning $\Lambda_{ref}$ have to obey the standard Dirac quantization conditions. So where are these additional states in the low energy effective field theory? The answer is that they are the additional
states associated with the SCFT itself. Indeed Dirac quantization and unimodularity are constraints for the tensorial Coulomb branch of the supergravity theory. As the SCFTs have their own Coulomb branches, these must be taken into account. In F-theory, the superconformal sectors come about from two-cycles in the base which have collapsed to zero size. The refined basis of string states comes from D3-branes wrapped over precisely these additional $\mathbb{P}^1$'s. The lattice $\Lambda_{ref}$ is then identified with the second homology lattice of the resolved base: that being a smooth compact oriented 4-manifold, the lattice is unimodular by Poincar\'e duality \cite{milnor}.

To further support this picture, we study a particular example of F-theory compactification on a Calabi-Yau threefold
with orbifold singularities in the base $B$. We take $B=\mathbb{P}^2/\mathbb{Z}_3$, and
show that the effective theories can be consistently
described by six-dimensional $\cn=1$ supergravity theories coupled to SCFTs.
When the elliptically fibered Calabi-Yau manifold $X \ra B$
is at a generic point in complex moduli space, the effective theory has three $A_2$ $(2,0)$ theories and 93
neutral hypermultiplets coupled to the $\cn=1$ supergravity theory.
By tuning the complex structure moduli, the effective theory can develop gauge symmetries.
We examine the various divisors a gauge symmetry can live on and describe the physics
in each case. As expected, when the gauge brane hits the orbifold locus, there exists a strongly coupled SCFT
living at the intersection that contributes to the gauge anomaly. We
verify that the anomaly cancellation conditions hold in each of these cases.

The rest of this paper is organized as follows. First, in section \ref{s:BSMOOVE}, we spell out the constraints
imposed by 6D anomaly cancellation and charge quantization both in 6D field theory terms, as well as in the
geometry of an F-theory compactification. In section \ref{s:BNOTSMOOVE} we turn to the case where the base
contains orbifold singularities, and therefore an SCFT sector. We show that anomaly cancellation can be understood in terms
of the intersection theory  of fractional divisors. Moreover, we explain how the conditions of charge quantization
are obeyed in this case. In section \ref{s:p2z3}, we study the case when
$B= \mathbb{P}^2/\mathbb{Z}_3$ in detail.
We conclude with comments and further directions of
research in section \ref{s:future}. Some technical details are collected in the appendices.

\section{F-theory on a smooth base \label{s:BSMOOVE}}

In this section we discuss some aspects of 6D supergravity theories for F-theory compactified on a \textit{smooth} base $B_{smth}$. Most of the material we shall review is well-known, and can be found in the existing literature.

%As far as we are aware, however, other results such as the proper charge quantization conditions for F-theory on a non-compact base do not seem to have been stated previously.

Recall that in F-theory, the type IIB axio-dilaton has a position dependent profile. To get an $\mathcal{N} = (1,0)$ theory in six dimensions, we work on the background $\mathbb{R}^{5,1} \times B_{smth}$ where $B_{smth}$ is the base of an elliptically fibered Calabi-Yau threefold, i.e.
%$T^2 \rightarrow X \rightarrow B_{smth}$.
$X \rightarrow B_{smth}$.
In minimal Weierstrass form, the defining equation for $X$ is:
\begin{equation}
y^2 = x^3 + fx + g
\end{equation}
where $f$ and $g$ are respectively sections of $\mathcal{O}_{B_{smth}}(-4 K_{B_{smth}})$ and $\mathcal{O}_{B_{smth}}(-6 K_{B_{smth}})$. The elliptic fiber may contain singularities, and the discriminant locus $4 f^3 + 27g^2 = 0$ tells us the locations of seven-branes wrapping curves in $B_{smth}$.

A hallmark of chiral 6D theories is the presence of self-dual and anti-self-dual three-form field strengths, and the corresponding BPS lattice of strings. These field strengths come about about from reduction of the 10D gravity multiplet, as well as reduction of the RR five-form flux to six-dimensional vacua. We get a single self-dual two-form potential $B_{\mu \nu}^{+}$ from the 6D gravity multiplet, and $T$ anti-self-dual two-form potentials $B_{\mu \nu}^{-}$ from the 6D tensor multiplets. Altogether, the corresponding two-form potentials rotate as a vector of $SO(1,T)$. We reach the tensorial Coulomb branch of the theory by giving vevs to the scalars in the tensor multiplets. This also generates a tension for the strings.

Anomaly cancellation via the Green-Schwarz-West-Sagnotti mechanism
\cite{Green:1984bx,Sagnotti1,Sadov:1996zm,anomalies}
dictates a delicate interplay between these tensor degrees of freedom,
and the vector multiplets of the 6D theory. For example, the invariant
field strengths of these tensor fields are given by
\be
H^M = d B^M + {1 \ov 2} a^M \om_{3L} +
\sum_i {2 b^M_i \ov \lambda_i} \om_{3Y}^i \,,
\ee
where the index $M$ runs from $0$ to $T$. The fields and parameters of the theory
can in fact be written in an $SO(1,T)$ invariant fashion, and the upper-case letters $M, N, \cdots$ are
used to denote these indices. %The tensor multiplets are then $SO(1,T)$ vectors.
The index $i$ labels the gauge group factors $G_i$ of the theory, while $\om_{3L}$ and
$\om_{3Y}^i$ are gravitational and gauge Chern-Simons three-forms, respectively.
The numerical factor $\lambda_i$, tabulated in table \ref{t:dynkin}, is the
Dynkin index for the fundamental representation of the gauge group $G_i$---it normalizes the
trace of the gauge group so that the minimum-charge instanton
has unit charge with respect to the fundamental
trace \cite{MR0459426,KMT2}.\footnote{We note that these differ from the
group theoretical factors used in \cite{OSTY} by a factor of two.}
We have assumed that the gauge group factors are all non-abelian, although incorporating
abelian group factors is straightforward \cite{Erler:1993zy}. %\footnote{ With a \emph{caveat}: if the non-abelian part of the gauge group has a non-vanishing third order Casimir, the term $F_{U(1)} \text{tr} F_G^3$ must cancel out separately.}
 The various multiplets of the effective theory,
which couple to the graviton and gauge fields,
have gravitational, mixed and gauge anomalies. Given that the total anomaly polynomial, which
is an eight-form $I_A$, factors in the form
\be
I_A = {1 \ov 32} \Om_{MN} X^M \wedge X^N
\label{factorization}
\ee
with the factors being the four-forms
\be
X^M = {1 \ov 2} a^M \tr R^2 +
\sum_i {2 b^M_i \ov \lambda_i} \tr F_i^2 \,,
\ee
the anomaly can
be cancelled by a local term
\be
\cL_{GS} = - {1 \ov 32} \Om_{MN} B^M \wedge X^N \,.
\ee
Here, $R$ and $F_i$ denote the Riemann
curvature and the $G_i$-field strength,
respectively. The trace ``$\tr$" without any index denotes
the trace taken with respect to the fundamental
representation. Given that the effective theory can be described
by using conventional $(1,0)$ supermultiplets
in six-dimensions, the total anomaly polynomial
$I_A$ can be computed by adding up the anomaly
polynomials of the individual multiplets, which we
summarize in appendix \ref{ap:anomaly polynomials}.

\begin{table}[t!]
\center
  \begin{tabular}{ | c || c | c | c | c | c | c | c | c | c |}
  \hline
 $G$ & $A_n$ & $B_n$ & $C_n$ & $D_n$ &
 $E_6$ &$E_7$ & $E_8$ & $F_4$ & $G_2$ \\ \hline\hline
 $\lambda$ & 1 &2& 1& 2& 6 &12& 60 &6 &2 \\ \hline
    \end{tabular}
  \caption{The Dynkin index for the fundamental representation
  of each group.}
\label{t:dynkin}
\end{table}

The symmetric matrix $\Om_{MN}$ of \eq{factorization} is a $SO(1,T)$ metric, which
can be understood as an integer-valued quadratic form on the string charge lattice
\cite{SeibergTaylor,Sagnotti1,FMS,SW6}. Geometrically, this
is just the intersection pairing on $H_{2}(B_{smth} , \mathbb{Z})$.
It is convenient to use this $SO(1,T)$ metric to raise and
lower indices, i.e.,
\be
\Om^{MN} \equiv (\Om^{-1})_{MN} \,.
\ee
The magnetic source $\tJ_M$ of the $M^{\text{th}}$
tensor field is given by
\be
\tJ^M =
dH^M = {1 \ov 2} a^M \tr R^2 +
\sum_i b^M_i \left( { 2 \ov \lambda_i} \tr F_i^2 \right) \,.
\label{magnetic}
\ee
Meanwhile, the self duality conditions of the
theory can be written as
\be
* \Om_{MN} H^N = G_{MN} H^N
\ee
where the elements of the matrix
$G_{MN}$ are given by
\be
G_{MN} = 2 j_M j_N - \Om_{MN} \,.
\ee
The star operator acts on differential forms
by the Hodge dual operation.
Here, the $SO(1,T)$ unit vector
$j^M$, i.e.,
\be
j^M j_M = \Om_{MN} j^M j^N =1 \,,
\ee
parametrizes the vacuum expectation
value of the $T$ scalars in the
$T$ tensor multiplets of the six-dimensional theory.
The electric source of the $M^{\text{th}}$ tensor
field is then given by \cite{Sagnotti1,FMS,SW6}
\be
J_M = d * G_{MN}H^N =
{1 \ov 2} a_M \tr R^2 +
\sum_i b_{i,M} \left( { 2 \ov \lambda_i} \tr F_i^2 \right) \,.
\label{electric}
\ee
We hence see that gauge instantons of the theory are electric/magnetic sources
for the tensor fields of the theory. The anomaly coefficients encode
the string charges of the BPS instantons---the tension of an instanton
with minimum charge is given by the inverse
gauge coupling
\be
b_i \cdot j \equiv \Om_{MN} b_{i}^M j^N \,.
\ee
The Dirac quantization condition for these strings impose that the vectors $b_i$ must be elements of an integral
lattice whose inner product matrix is given by $\Om$.

For F-theory models on a smooth compact base $B_{smth}$, we have $H_{2}(B_{smth} , \mathbb{Z}) = \Lambda_{string}$,
the string charge lattice. The matrix $\Om_{MN}$ is then the intersection pairing matrix of
the homology cycles---this lattice is integral and unimodular, due to
Poincar\'e duality. Then, $a$ and $b_i$ have the geometric interpretation
as being the homology classes of the canonical divisor and the divisor
the $G_i$ seven-brane wraps \cite{Sadov:1996zm,KMT1,KMT2}. The vector $j$ can then be understood as the
K\"ahler class of the base manifold.

\subsection{Charge quantization and unimodularity}

Let us say a few more words on the charge quantization conditions and their geometric avatars in an F-theory compactification.
Geometrically, the lattice of string charges $\Lambda_{string}$ is simply the homology lattice for the compact model $B_{smth}$, i.e. we have:
\begin{equation}
\Lambda_{string} = H_{2}(B_{smth} , \mathbb{Z}).
\end{equation}
As we have remarked above $H_{2}(B_{smth} , \mathbb{Z})$ is automatically unimodular. This fact can also be understood purely in terms of the 6D supergravity theory \cite{SeibergTaylor} (see also \cite{BanksSeiberg,HellermanSharpe}). Along the tensorial Coulomb branch, Dirac quantization in flat $\mathbb{R}^{5,1}$ implies that the allowed string charges have to be integer valued \cite{Deser:1997se}.

Additional constraints can follow from studying a 6D effective theory on different backgrounds. In particular, the existence of a partition function imposes the condition that the lattice of string charges is in fact unimodular \cite{SeibergTaylor}. As explained in \cite{SeibergTaylor}, to establish this condition, consider the 6D theory on a $\mathbb{R}^{1,1}\times \mathbb{CP}^2$ background to obtain a chiral 2D theory whose charge lattice is identified with the string charge lattice. Now, consider the partition function of such 2D theory on a torus $T^2 = S^1_a \times S^1_b$. The $S$ transformation that exchanges $S^1_a$ with $S^1_b$ is always a symmetry of the theory, and therefore the partition function has to be invariant with respect to it. As explained in \cite{SeibergTaylor}, the $S$--invariance of the 2D partition function is realized only if the charge lattice of the 2D model is self--dual, i.e. unimodular. This implies that the unimodularity of the string charge lattice in 6D is a necessary condition for the theory to have a well-defined partition function on $\mathbb{CP}^2 \times T^2$, and therefore a necessary condition to have a consistent supergravity theory. This is why the string charge lattice of the tensor fields in a consistent 6D supergravity theory must be unimodular.

Of course, some well-known 6D theories do not satisfy this condition of a unimodular intersection form. A notable class of examples are the $(2,0)$ theories of $A_N$ type. Indeed, we must note that a theory of (anti-)self-dual fields that does not have a well-defined partition function on a manifold nevertheless can define a sensible quantum field theory \cite{Witten:1996hc,Aharony:1998qu,Witten:1998wy,Moore:2004jv,Belov:2006jd,Freed:2006yc,Witten:2007ct,Witten:2009at,Henningson:2010rc,Freed:2012bs}.  Such a theory, referred to as a ``relative quantum field theory" in \cite{Freed:2012bs}, has a partition bundle (or a partition vector) over the geometric moduli space of the manifold as opposed to a partition function: additional topological data must be specified to fully characterize the behavior of the model in curved spacetimes. These theories, however, on their own cannot be coupled to gravity in a consistent way, as they can be thought of as having anomalies under large diffeomorphisms. In a theory coupled to gravity, however, there is no such issue since it is automatically diffeomorphism invariant, and in particular has a unimodular lattice of string charges.\footnote{We thank Y. Tachikawa and W. Taylor for explaining these points to us.} Finally, let us remark that even if a 6D field theory cannot be consistently coupled to 6D supergravity, it is nevertheless always possible to embed the lattice of strings in a unimodular lattice.

\section{The case of an orbifold base \label{s:BNOTSMOOVE}}

Having reviewed the case of F-theory on a smooth base, we now turn to the study of F-theory on a base with orbifold singularities. Roughly speaking, the physical picture is that our base will now contain various ``fractional divisors'' which can be wrapped by seven-branes, as well as D3-branes. Geometrically, these fractional divisors will pass through the locus of the orbifold singularity. As such, it is important to understand whether we can still make sense of the resulting theory. The seven-branes will contribute gauge theory sectors, and the D3-branes will contribute BPS strings with tension. Owing to the fact that there is a singularity in the base, we can also expect there to be additional light states which contribute to the low energy effective theory. These states are the contribution from a 6D SCFT. The F-theory geometry provides a systematic way to couple these systems to gravity.

Remarkably, many aspects of the 6D effective theory can be understood purely in terms of the geometry of these fractional divisors. For example, we find that the anomaly polynomial for such theories can be understood purely in terms of the intersection theory of $H_{2}(B_{orb} , \mathbb{Z})$. On the other hand, we will also see that charge quantization \textit{predicts} the existence of additional light states in the low energy effective field theory. These states are simply the contributions from the `internal' degrees of freedom of the SCFTs.

\subsection{Geometric preliminaries}

Since it will form the core of our mathematical analysis, we first review some salient features of intersection theory on an orbifold base $B_{orb}$. The key point is that the intersection numbers for cycles in $H_{2}(B_{orb} , \mathbb{Z})$ will be rational numbers. %In the physical theory, charge quantization will then predict the existence of additional states.
To avoid cluttering the notation, we shall drop the ``orb'' from $B_{orb}$ in what follows.

To begin, we shall always assume the existence of a smooth resolution $\hat B \rightarrow B$.
Denote by $e_M$ for $M = 0,..., T$ the basis of divisors for $\hat B$. To reach the orbifold point,
we shall blowdown some subset of these divisors. Denote this collection by:
\be
D_m = D_m^M e_M, \quad m=T_0+1,\cdots,T \,.
\ee
The orbifold point is reached by tuning the K\"ahler class
$j$ such that
\be
j \cdot D_m = 0\,.
\ee
Viewed as a vector in $H_2({\hat B} , \mathbb{R})$, $j$ is
thus restricted to lie in the orthogonal
complement of the subspace $V_S$\
spanned by $\{D_m \}$
with respect to the inner product space
$H_2(\hat B,\mathbb{R})$.
%\footnote{Note that we have used the notation $V_S$,
%as opposed to $\Lambda_S$, as $D_m$
%is not guaranteed to be an integral basis
%for the entire lattice:  $\Lambda_S \equiv \Lambda \cap V_S $.}
We identify this orthogonal complement with the
inner product space $H_2(B,\mathbb{R})$
of the surface $B$ obtained by blowing down
the divisors $D_m$. It is convenient to take an integral basis
\be
u_{\mu}, \quad \mu=0,\cdots,T_0
\ee
of the $SO(1,T_0)$ sublattice $\Lambda_0 = V_S^{\perp} \cap \Lambda$
of $\Lambda$. By definition,
\be
u_{\mu} \cdot D_m =0
\ee
for any $m$ and $\mu$.
We consistently use the labels $m, n$ (resp. $\mu, \nu$)
to label indices in the range $\{T_0+1,\cdots, T \}$
(resp. $\{0,\cdots,T_0 \}$), respectively.
Taking $u_\mu$ and $D_m$ to be the new
basis for $H_2 (\hat B, \mathbb{R})$,
the intersection matrix now factors into the form:
\be
\Om'_{M'N'} = \Om^0_{\mu \nu} \oplus \Om^S_{mn} \,,
\label{orth}
\ee
with
\be
\Om^0_{\mu \nu} \equiv u_\mu \cdot u_\nu,\quad
\Om^S_{mn} \equiv D_m \cdot D_n \,.
\ee
As before, we raise and lower the $\mu, \nu$ (resp. $m, n$)
indices using the metric $\Om^0_{\mu \nu}$ (resp. $\Om^S_{mn}$).

Though it is tempting to identify $\Lambda_0$ as the integral homology lattice of the orbifold $B$, this is not quite true. Letting $V_0$ to be the inner product space spanned by $u_\mu$ over the reals, the integral homology lattice $H_2(B,\bZ)$ of $B$ is given by the orthogonal projection of the homology lattice $\Lambda = H_{2}({\hat B} , \mathbb{Z})$ of its resolution to $V_0$.
Due to the unimodularity of $\Lambda$, the homology lattice of $B$ can be shown to be
given by the dual of $\Lambda_0$:
\be
\Lambda_0^* = \{ \ell  \in V_0 \, : \,
\ell \cdot \ell' \in \bZ ,~\text{for all}~
\ell' \in \Lambda_0 \} \,.
\ee
Since $B$ is an orbifold, as opposed to being a smooth manifold, $\Lambda_0^*$ is strictly larger than $\Lambda_0$. Hence, the lattice $\Lambda_0$ is not unimodular and $\Lambda_0^*$ is not integral, but rational.

An equivalent, algebraic definition of $H_2(B,\bZ)$
can be given \cite{BG}, since $B$ can be treated
as a rational surface ($H^{2,0} (B,\bC) = 0$). $H_2(B,\bZ)$ is the group of divisors of $B$ modulo
algebraic equivalence---with suitable definitions
of divisors and algebraic equivalence for
orbifolds. A divisor on an orbifold can be $\bQ$-Cartier,
but not Cartier%
---that is, on its own, it may not have a good defining
equation, while a multiple of it has one.
$\bQ$-Cartier divisors can be identified
as the divisors that are Weil, but not Cartier. These divisors,
which we shall often refer to as ``fractional divisors,"
can have fractional intersection numbers
with other divisors.

An operational definition of Weil and Cartier divisors
can be given by the following: Weil divisors can be
understood as divisors that have a well defined locus,
while Cartier divisors are divisors whose defining equation,
in each patch of the algebraic variety, lies in the
ring of rational functions of that patch. The homology class of the
Cartier divisors are elements of the integral lattice $\Lambda_0$,
while the homology class of the fractional divisors
are elements of $\Lambda_0^*$ that do not
lie on the integral lattice points. A more rigorous treatment of divisors
of complex orbifolds can be found in section 4.4 of \cite{BG}.

There is an intuitive way of describing the origin of
the fractional divisors from the point of view of the
homology lattice. The map of the homology class of a divisor
on $\hat B$ to $B$, upon the birational map
of blowing down the divisors $D_m$,
is given by the projection from the homology lattice
$\Lambda$ to $\Lambda_0^*$.
In particular, the canonical class of $\hat B$
maps to that of $B$ in this way.%
\footnote{See, for example, proposition 4.4.15 and equation (4.4.2) of \cite{BG}.}
Since $\Lambda_0$ is not unimodular,
there exist integral divisors $D$ of $\hat B$
that have fractional coefficients
when written as a linear combination of the basis
$\{ u_\mu \} \coprod \{ D_m \}$.
This is because the unimodularity of $\Lambda_0$ and the requirement that the projection of any
lattice vector in $\Lambda$ to $V_0$
lies in the integral lattice in $\Lambda_0$ are equivalent facts.%
\footnote{ Proof: the projection $v_0$
of a vector $v$ in $\Lambda$ to $V_0$ is given by
\be
v_0 = (v \cdot u^\mu) u_\mu \,.
\ee
If $\Lambda_0$ is unimodular,
$v_0$ obviously lies within $\Lambda_0$,
as $u^\mu$ is an integral vector in $\Lambda$.
Meanwhile, if $v \cdot u^\mu$ is integral
for any $v$ and $\mu$, $u^\mu$ itself is an integral
vector, which lies in $\Lambda_0$. Hence,
$\Lambda_0$ must be unimodular.}
Such divisors become fractional
upon projecting down to $\Lambda_0^*$,
i.e., when $\hat B$ is blown down to $B$.

\subsection{Anomalies and fractional divisors}\label{fracanoX}

Having dispensed with the geometric preliminaries, we now turn to the study of 6D supergravity theories coupled to SCFTs. The big surprise is that the anomaly polynomial for these theories can be recast purely in terms of the intersection
theoretic data on $H_2(B_{orb}, \mathbb{Z})$ alone. In other words, the SCFTs constitute a ``black box'' which effectively generalizes
the case of contributions from more conventional matter fields such as 6D hypermultiplets.

Our starting point will be F-theory compactified on a smooth base ${\hat B}$ which degenerates to a base $B$
that contains orbifold singularities. Denote by ${\hat X}$ the corresponding elliptic Calabi-Yau with base ${\hat B}$.
We use the same notation as in the previous subsection, e.g. we let $\Lambda = H_{2}({\hat B} , \mathbb{Z})$ denote the lattice of BPS strings for the smooth phase. We move to the case with orbifold singularities
by collapsing a subset of the divisors of $\hat B$:
\be
D_m = D_m^M e_M, \quad m=T_0+1,\cdots,T \,.
\ee
We shall be interested in studying the physical theory defined by the divisors which remain at finite volume. Seven-branes
wrapping such fractional divisors will support vector multiplets, and D3-branes wrapped over such fractional divisors
correspond to strings. With the same notation of the previous section, in the new basis $j$ can be written as
\be
j = u_{\mu} j^{\mu} \,,
\ee
with
\be\label{projjX}
j^{\mu} = (u^\mu \cdot e_M) j^M \,.
\ee
The $(T_0+1)$ tensor fields $\cB^{\mu}$ which are not part of the SCFT degrees of freedom are also
aligned along the subspace of $H_2 (\hat B, \bR)$ spanned by
$u_{\mu}$, i.e., the inner product space $H_2(B, \bR)$.
$\cB^{\mu}$ is related to $B^M$ by
\be
\cB^{\mu} = (u^\mu \cdot e_M) B^M \,.
\label{0tensors}
\ee
Meanwhile, the tensor fields $\cB_S^m$, whose electrically
charged strings become tensionless can be identified as
\be
\cB_S^m \equiv (D^m \cdot e_M) B^M \,.
\label{stensors}
\ee
These tensors are part of the SCFT.
The corresponding gauge invariant field strengths of the
tensors are denoted by $\cH^\mu$ and $\cH^m_S$.
A consistency check that the tensors that are not part of the conformal subsector should be identified as \eq{0tensors} is to observe that
under this identification,
none of the tensionless strings of the SCFT carry
electric charge under $\cB^{\mu}$. Indeed the electric string current four-form is given by
\be
J_{\mu} =
d * (2 j_{\mu} j_\nu - \Om^0_{\mu \nu} ) \cH^\nu \,,
\label{electric}
\ee
and our claim follow from the orthogonal splitting \eq{orth} of the lattice.\footnote{We remind the reader that the indices $\mu, \nu$
of equation \eq{electric} are the $SO(1,T_0)$ indices---the
space-time indices in this equation are suppressed.}

The (local) gauge group of the theory
on $\hat B$ can be factored into two pieces,
\be
\cG = \cG_0 \times \cG_S
= \prod_k G_{0,k} \times \prod_\kappa G_{S,\kappa} \,,
\ee
where the second factor denotes the gauge groups which become
strongly coupled.
The seven-branes responsible for the gauge
symmetry wrap a linear combination of cycles
that are being blown down.
Hence, the gauge anomaly
coefficients of the gauge groups $G_{S,\kappa}$
are linear combinations only of the shrinking cycles $D_m$,
\be
b_{S,\kappa} = b_{S,\kappa}^{m} D_m \,,
\ee
where the coefficients $b_{S,\kappa}^m$ are all
integral.
Notice that from equation \eq{projjX} it follows that
\be
j \cdot b_{S,\kappa} =0 \,,
\ee
so that the instantons (i.e. the strings) of the gauge group $\cG_S$
become tensionless.
Meanwhile, the anomaly coefficients of $G_{0,k}$
and the gravitational anomaly coefficient
can in general have components in the $D_m$ directions.
We can decompose them in the following way:
\bea
a &=   a^\mu u_\mu + a^{m} D_m  \equiv a_0 + a_\cS \,, \\
b_{k} &=  b_{k}^\mu u_\mu + b_{k}^{m} D_m \equiv b_{0,k} + b_{\cS,k} \,.
\label{projected coefficients}
\eea
$a_0$ and $b_{0,k}$ are projections of the
coefficients $a$ and $b_k$ to $H_2(B,\bR)$.

The Green-Schwarz term of the effective theory
on $\hat X$ now can be decomposed as
\bea
\cL_{GS} &= -{1 \ov 32} \Om_{\mu \nu} \cB^\mu X^\nu
-{1 \ov 32} \Om_{m n} \cB_S^{m} X^n
\equiv \cL_0 + \cL_{S} \,,
\label{GStot}
\eea
where
\bea
X^\mu &= d \cH^\mu =
{1 \ov 2} a_0^\mu \tr R^2 +
\sum_k b^\mu_{0,k} \left( { 2 \ov \lambda_{0,k}} \tr F_{0,k}^2 \right) \\
X^m &= d \cH_S^m =
{1 \ov 2} a_\cS^m \tr R^2 +
\sum_k b^m_{\cS,k} \left( { 2 \ov \lambda_{0,k}} \tr F_{0,k}^2 \right)
+\sum_\kappa b^m_{S,\kappa}
\left( { 2 \ov \lambda_{S,k}} \tr F_{S,\kappa}^2 \right) \,.
\label{magnetic decomp}
\eea
Recall that the effective theory on the tensor
branch of the superconformal theory is a $(1,0)$ field theory
with the tensor multiplets $\cB^m_S$,
the gauge group $\cG_S$, hypermultiplets
charged under $\cG_S$ (some of which can
carry charge under $\cG_0$),
and in certain cases,
some neutral hypermultiplets
\cite{HMV,DHTV,Heckman,OST,OSTY,Intriligator}.
Let us denote the
one-loop anomaly polynomial of the fields used
to describe the SCFT,
upon coupling its stress energy
tensor and flavor currents to a background graviton
and gauge fields, as $I_{S,1\ell}$.
The total anomaly polynomial of the low energy
supergravity theory on $\hat X$ then decomposes into
\be
I_\text{tot} = I_{0,1\ell} + I_{S,1\ell} \,.
\ee
The one-loop anomalies $I_{0,1\ell}$
come from the supergravity multiplet,
the $T_0$ tensor multiplets that are not part of the
SCFT, gauge multiplets of $\cG_0$
and hypermultiplets that are either neutral
or carry charge only under $\cG_0$.
Assuming that the low energy supergravity
description of the F-theory compactification
on $\hat X$ is consistent, we find that the
anomaly cancellation condition
\be
I_{0,1\ell} + I_{S,1\ell} =
{1 \ov 32} \Om_{\mu \nu} X^\mu X^\nu
+{1 \ov 32} \Om_{m n} X^{m} X^n
\label{anomaly cancellation}
\ee
is satisfied.

Now the piece $\cL_S$ of \eq{GStot}
is precisely the Green-Schwarz term
of the effective theory of the superconformal theory on
the tensor branch, as it is the piece that
only involves the tensors $\cB_S^m$.
Then the anomaly polynomial of the SCFT coupled
to supergravity and gauge fields in gauge group
$\cG_0$ is given by \cite{OSTY,Intriligator}
\be
I_S = I_{S,1\ell} -{1 \ov 32} \Om_{mn} X^m X^n \,.
\label{IS}
\ee
A simple consistency check of the fact that
$\cL_S$ is the correct Green-Schwarz term is
that, due to \eq{anomaly cancellation},
all the gauge and mixed anomalies involving
the gauge symmetry group $\cG_S$ are cancelled
in \eq{IS}---the only remaining terms in $I_S$
involve the metric curvature and gauge field
strengths of $\cG_0$.

The F-theory compactification
on the singular base $X \ra B$ leads to an effective theory that can be described by
a supergravity theory with $T_0$ tensor
multiplets and gauge symmetry $\cG_0$
interacting with a strongly coupled superconformal system \cite{DHTV,Heckman}.
Let us now show that the anomalies of this
theory are cancelled by the GSSW mechanism
with anomaly coefficient vectors $a_0$ and $b_{0,k}$.
Note that in this effective theory, the Green-Schwarz
term now becomes $\cL_0$ of \eq{GStot},
which only involve the tensor fields
$\cB^\mu$.
The total anomaly polynomial of the theory
is given by
\be
I_{0,\text{tot}}=I_{0,1\ell} + I_S \,.
\ee
The first term is the contribution of the conventional
fields of the supergravity theory, while the second
term comes from the strongly coupled sector.
Due to the computation \eq{IS}, and the anomaly
cancellation condition \eq{anomaly cancellation}
on the compactification on $\hat X$,
we find that
\be
I_{0,\text{tot}} = {1 \ov 32} \Om_{\mu \nu} X^\mu X^\nu \,,
\ee
which is precisely cancelled by the Green-Schwarz term
$\cL_0$!

We also see that the gravitational and gauge
anomaly coefficients in this equation are given by
$a_0$ and $b_{0,k}$ defined in \eq{projected coefficients}.
Notice that this does not require us to pass onto the
tensor branch of the 6D SCFT. Indeed, upon passing onto the tensor branch,
we get additional gauge theory sectors, but none of the instantons of the gauge
group $\cG_S$ are charged under the tensor fields
$\cB^\mu$ of this theory.

The discussion above implies that the anomaly
coefficients $a_0$ and $b_{0,k}$ of the effective
theory of F-theory compactified on $X$ still have
the geometric interpretation as respectively specifying the homology class
of the canonical class
and the seven-brane loci. However, if a seven-brane carrying gauge group $G$
is wrapping a fractional
divisor $\beta$ of $B$ whose homology class is
given by $b = [\beta]$,
intersection numbers involving $b$,
e.g.,
\be
\Om_{\mu \nu} b^\mu b^\nu \,,
\label{frac anom}
\ee
can be fractional. Since $\beta$ comes from the
projection of an integral divisor $\hat \beta$
in $\hat B$,
$\hat b = [\hat \beta]$ must have
components lying along the directions of the
cycles blown down. In fact, a fractional divisor
$\beta$ intersects the orbifold loci where
the resolution divisors contained in $\hat \beta$
are localized.
Physically, this implies that
the $G$-instantons are charged under
tensor fields that are part of the strongly coupled
subsector. The anomaly polynomial
$I_S$ of the superconformal theory has the proper fractional coefficients
 to offset the fractionality of the intersection
number \eq{frac anom}. However,
this also implies that the string charge lattice of
an F-theory background with a superconformal sector
cannot be identified with the homology
lattice of the singular base $B$, as this would violate the quantization of charges over the integers---it must be identified with the homology lattice of the base $\hat B$ obtained by
resolving all the strongly coupled singularities, as we are going to argue in the following section.

Let us point out that $\beta$ can be a Cartier divisor
on $B$ and its blow-up $\hat \beta$ still can have
components of $D_m$ in it. When $\beta$ is a Cartier
divisor that does not intersect any orbifold points, its
homology class remains in the sublattice $\Lambda_0$
of $\Lambda$ even after blowing up $B$ into $\hat B$.
On the other hand, when $\beta$ intersects the orbifold
point, $\hat \beta$ has components that are orthogonal
to $\Lambda_0$ within $\Lambda$. Physically, this
is because the seven-brane intersects the locus where
an SCFT resides, and hence its instantons carry charge
under the tensor fields of that SCFT.

\subsection{Charge quantization and unimodularity}

The results of the previous subsection are perhaps surprising. Without needing to specify \textit{any} details
of the microsopic theory generated by our SCFT, the coarse data of the
homology lattice $\Lambda_0^* = H_{2}(B_{orb} , \mathbb{Z})$ is sufficient to extract the details of the
anomaly polynomial for the 6D theory. In this sense, one might loosely refer to $\Lambda_0^*$ as the ``anomaly lattice'', since this suffices to fix these properties of the macroscopic theory.

However, $\Lambda_0^*$ cannot be interpreted as the string charge lattice of the model, because the condition of charge quantization and unimodularity would be clearly violated. Recall that for a 6D field theory on the tensorial Coulomb branch, Dirac quantization imposes the condition that there is an integer valued pairing for the complete Hilbert space of states. Indeed, superconformal systems have their own tensor multiplets, and the charge quantization and unimodularity constraints are expected to hold only when \emph{all} scalars belonging to tensor multiplets are given vevs. The fact that $\Lambda_0^*$  cannot be interpreted as the string charge lattice of the model is not so surprising: such lattice has to include all strings coming from the SCFTs as well. In the geometry, this is the requirement that the intersection form on the resolved geometry ${\hat B}$ is valued in the integers. Said differently, the resolution $\hat B$ provides a refinement of the lattice of fractional divisors:
\begin{equation}
H_{2}(B , \mathbb{Z}) \subset H_{2}({\hat B} , \mathbb{Z})
\end{equation}
or equivalently:
\begin{equation}
\Lambda_0^*\subset \Lambda_{string}.
\end{equation}
Now, upon projecting $\hat B$ to the homology lattice of $B$, we lose some data. That is, if we have
two seven-branes wrapping different divisors in $\hat B$, the image under the projection map may be identical. Nevertheless,
the gauge instantons can have inequivalent charges with respect to the
tensor multiplets that are part of the superconformal theory.
The homology class of a divisor $\beta$ of $B$
is only \textit{part} of the information that specifies $\beta$: a divisor is an algebro--geometric object rather than a topological
one. In particular, by knowing $\beta$, we can also
compute the homology class of the divisor $\hat \beta$
obtained from $\beta$ upon the resolution of $B$ to
$\hat B$. When a seven-brane with gauge group $G$ wraps
the divisor $\beta$ of $B$ the anomaly coefficient of the
gauge group $G$ is identified with $[\beta] \in \Lambda_0^*$. However,
the string charge of a unit $G$-instanton is given by
$[\hat \beta] \in \Lambda$.

This ``loss of data'' is really a hallmark of
having an SCFT coupled to gravity, and can occur even when there is
no orbifold singularity present in the base. For example, the
superconformal matter of \cite{DHTV, Heckman} comes about
when we have a collision of two components of the discriminant locus, at which point the elliptic fiber
becomes too singular to satisfy the Calabi-Yau condition. Introducing the requisite blowups at this collision point, we obtain additional
curves in the base geometry. These additional curves produce a refinement of the original lattice of BPS charges. Observe, however, that by construction, the blowdown of these extra curves leads us back to a smooth base.

\subsection{Example: The $\mathcal{T}_{p}(N,M)$ theories}

To illustrate some of the above points, we will now turn to
some explicit non-compact examples. We shall couple these examples to gravity
in section \ref{s:p2z3}. We introduce the theories
$\mathcal{T}_{p}(N,M)$ which are defined by intersecting two
collections of non-compact seven-branes with respective gauge groups $SU(N)$ and
$SU(M)$ at a point in the base with an $A_{p-1}$
singularity, which is a $\mathbb{Z}_{p}$ orbifold singularity of the
form $\mathbb{C}^2 / \mathbb{Z}_{p}$, where the groups acts on the holomorphic coordinates $(u,v)$ of $\mathbb{C}^2$
as $(u,v)\mapsto (\omega u,\omega^{-1}v)$, where $\omega$ is a primitive $p^{\text{th}}$ root of unity.

In the terminology of \cite{DHTV}, the $\mathcal{T}_{p}(N,N)$ can be indentified with $T(SU(N),p-1)$ theories, while the $\mathcal{T}_{p}(N,M)$ with $N\neq M$ are examples of $T(SU(N),p-1)$ theories with decorations by T-brane data. Equivalently, these are engineered in Type IIA with a non-zero Romans mass---using Nahm pole boundary conditions for D8-D6-NS5 systems \cite{GaTom}.

An important feature of these theories is that although they involve the collision of two non-compact seven-branes, the ``matter'' living at the intersection point is itself a strongly coupled superconformal theory. In other words, this is an example of a ``superconformal matter'' system in the sense of reference \cite{DHTV, Heckman}. As such, they are also an excellent test case for studying the structure of 6D anomaly cancellation and charge quantization.

Let us give more details on the geometric realization of these theories. We consider F-theory on the base
$B_{orb} = \mathbb{C}^{2} / \mathbb{Z}_{p}$. Since this base is already Calabi-Yau, we can actually consider a trivial fibration. Then,
we get the $A_{p-1}$ type $(2,0)$ theory. When the elliptic fibration is non-trivial and contains non-abelian seven-branes
we get a $(1,0)$ SCFT with additional seven-branes in the geometry. These seven-branes can either wrap the compact cycles obtained by resolving the orbifold singularity, or can also correspond to non-compact divisors. In fact, 6D anomaly cancellation usually
correlates these contributions. As we will see shortly, these theories turn out to
have fractional anomaly coefficients, quantized in units of $p^{-1}$. This is compatible with the fact
that the intersection numbers of Weil divisors on a surface with an $A_{p-1}$ singularity can have
fractional intersection numbers in units of $p^{-1}$.

For expository purposes, we focus on the case where there is an $SU(N)$ seven-brane supported on a non-compact divisor $u = 0$, and an $SU(N + p k)$ seven-brane supported on another non-compact divisor $v = 0$. These seven-branes pass through the orbifold fixed point, and so to properly cancel all 6D anomalies, we can expect additional light degrees of freedom to be present. As a point of notation, let $\tB$ denote the covering space for $B_{orb} = \mathbb{C}^{2} / \mathbb{Z}_{p}$. Now let us consider the divisor $D_U$, defined by the equation $u =0$. The locus of
the divisor $D_U$ is well defined---the locus
$u=0$ on the covering manifold $\tB$ is $\Gamma$-%
invariant. The polynomial defining the divisor, $u$, however,
is not $\Gamma$-invariant. In more mathematical terms,
it does not lie in the ring of rational functions of $B$.
The same goes for the divisor $D_V$, defined by the equation $v=0$.
These divisors are ``fractional"---they are Weil, but not Cartier.
The Cartier divisors of $B$
are defined by elements of the
ring of rational functions of $B$, which is generated
by the $\Gamma$-invariant combinations
\be
u^p, \quad v^p , \quad \text{and} \quad uv
\ee
of $u$ and $v$.

Now let us consider an elliptic fibration over
$B$ with an $A_N$ singularity along the divisor
$U$. Writing the local Weierstrass model for the
elliptic fibration, we see that when
$N$ is not divisible by $p$, an additional
singularity must be present along the divisor $D_V$.
That is, using the local coordinates of the cover
$\tB$, we see that
in order for the Weierstrass model for
the fibration to be $\Gamma$-invariant, it must
be of the form
\be
xy = u^{N} v^{N+pk} \,,
\ee
for some integer $k$ such that $N+pk$ is
non-negative. The SCFT lying at the orbifold
point then has $SU(N) \times SU(N+pk)$
global symmetry, and we denote this theory as $\thy{N}{k}{p}$.

\begin{figure}[!t]
\centering\includegraphics[width=11cm]{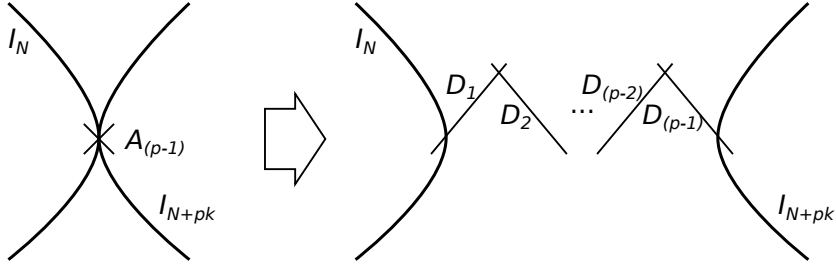}
\caption{\small The tensor branch of
the SCFT $\thy{N}{k}{p}$. On the left,
$\thy{N}{k}{p}$ is localized at the $A_{p-1}$ locus
which two flavor branes,
each of type $I_N$ and $I_{N+pk}$, pass through.
The $A_{p-1}$ singularity in the base
is resolved on the right, by introducing the
resolution divisors $D_m$, hence moving on
to the tensor branch of the theory.
A singular fiber of type $I_{N+mk}$ fibers over
the divisor $D_m$. The effective gauge group
of the tensor branch theory can thus be identified
as $\prod_{m=1}^{p-1} SU(N+mk)$.}
\label{f:TNkp}
\end{figure}

The SCFT can be taken to a generic
point in its tensor branch by blowing
up the $A_{p-1}$ singularity, and arriving
at the non-singular manifold $\hat B$.
The resolution divisors
\be
D_1, \cdots, D_{p-1}
\label{res divisors}
\ee
are $(-2)$ curves whose intersection
matrix is given (up to an overall minus sign) by the Cartan matrix
of $SU(p)$. Let $D_1$ be the divisor
adjacent to the $SU(N)$ locus and $D_{p-1}$
be adjacent to the $SU(N+pk)$ divisor in
the resolved manifold.
The effective theory of the SCFT
on the tensor branch is a theory with
$(p-1)$ tensor fields with
\be
SU(N+k) \times SU(N+2k) \times
\cdots \times SU(N+(p-1)k)
\ee
gauge symmetry, where for each pair of
adjacent gauge groups, there exists
a bifundamental hypermultiplet.
There are also $N$ hypermultiplets of
$SU(N+k)$ that can be thought of as
a bifundamental
between $SU(N+k)$ and the $SU(N)$ flavor group,
and $N+pk$ hypermultiplets of
$SU(N+(p-1)k)$ that can be thought in
an analogous way. The geometry of the
resolved singularity is shown in figure \ref{f:TNkp}.
The one-loop contribution of the effective fields
to the total anomaly polynomial is then given by
\bea
I_{S,1 \ell}
&= {1 \ov 5760} \left( 30p-30+N^2 + Nkp +{p(p-1) \ov 2} k^2 \right)
\left(\tr R^4 + {5 \ov 4} \tr R^2 \right)\\
&-{(p-1) \ov 128} (\tr R^2)^2
-{1 \ov 4} \sum_{m=1}^{p-1}
 (\tr F_m^2)^2
+ {1 \ov 4} \sum_{m=1}^{p-2} \tr F_m^2 \tr F_{m+1}^2\\
&+{1 \ov 4} \tr F_0^2 \tr F_{1}^2
+{1 \ov 4} \tr F_{p-1}^2 \tr F_{p}^2 \\
&-{1 \ov 96} (N+k) \tr R^2 \tr F_0^2
-{1 \ov 96} (N+(p-1)k) \tr R^2 \tr F_p^2 \\
&+{1 \ov 24} (N+k) \tr F_0^4 +{1 \ov 24} (N+(p-1)k) \tr F_p^4 \,,
\label{Nkp 1loop}
\eea
where the first term comes from counting all
the effective fields:
\be
29 (p-1) - \sum_{m=1}^{p-1} ((N+mk)^2-1)
+ \sum_{m=0}^{p-1} (N+mk)(N+mk+k) \,.
\ee
We have used $F_m$ to denote
the field strength of the $SU(N+m k)$ gauge group.
We have also included the anomalies of the flavor symmetries,
whose background field strengths are denoted by
$F_0$ and $F_p$.

Taking the basis of the
homology lattice to be the divisors
\eq{res divisors},
the anomaly coefficients
of the theory are given by
\be
D_m X^m =
2\hat D^1 \tr F_0^2 +
2\hat D^{p-1} \tr F_p^2 +
\sum_{m=1}^{p-1} 2 D_m \tr F_m^2 \,.
\label{anom coeff Nkp}
\ee
We have also included the anomaly coefficients
for the flavor symmetries.
%Note that since we are dealing with a theory with gravity decoupled (our base is non-compact)
%there is no gravitational anomaly to consider.
Note that there is no $\tr R^2$ term in \eq{anom coeff Nkp}.
{This is because the canonical class of the base
manifold is trivial, as it is a Calabi-Yau resolution of
an $A_{p-1}$ singularity.
Recall that the coefficient of the $\tr R^2$ term in
the four-form $D_m X^m$ is given by the projection of
the canonical class of the base to the divisors $D_m$
\eq{magnetic decomp}. The coefficient of $\tr R^2$
obtained this way only depends on the local geometry
of the resolution, and does not depend on the global
embedding of the singularity.}
The divisor $\hat D^m$ denotes the
dual divisor of $D_m$:
\be
\hat D^m \cdot D_n = \delta^m_n \,.
\ee
$\hat D^1$ and $\hat D^{p-1}$ can
be explicitly written as
\bea
\hat D^1 = {1 \ov p} \sum_{m} {(m-p)} D_m, \qquad
\hat D^{p-1} = -{1 \ov p} \sum_{m} {m} D_m \,.
\eea
Not coincidentally, the choice for the flavor
anomaly coefficients in \eq{anom coeff Nkp}
is the unique choice that cancels the one-loop
anomalies involving matter jointly charged under the
flavor and gauge symmetries---i.e., the third line of
\eq{Nkp 1loop}---consistent with the
picture of \cite{OSTY, Intriligator}.
Geometrically, $\hat D^1$ and $\hat D^p$ have
the interpretation as the projection of the flavor
brane locus on the resolved manifold $\hat B$
to the compact basis $D_m$.
The total anomaly of the theory, then, is given by
\bea
I_S &= I_{S,1\ell} - {1 \ov 32} (D_m \cdot D_n) X^m X^n \\
&= {1 \ov 5760} \left( 30p-30+N^2 + Nkp +{p(p-1) \ov 2} k^2 \right)
\left(\tr R^4 + {5 \ov 4} \tr R^2 \right)\\
&-{(p-1) \ov 128} (\tr R^2)^2
-{1 \ov 96} (N+k) \tr R^2 \tr F_0^2
-{1 \ov 96} (N+(p-1)k) \tr R^2 \tr F_p^2 \\
&+\left[ {1 \ov 24} (N+k) \tr F_0^4 +{1 \ov 8} {(p-1) \ov p}(\tr F_0^2)^2 \right]  \\
&+\left[ {1 \ov 24} (N+(p-1)k) \tr F_p^4 +{1 \ov 8} {(p-1) \ov p} (\tr F_p^2)^2 \right]
+ {1 \ov 4 p} \tr F_0^2  \tr F_p^2
\,.
\label{anom Nkp}
\eea
Here, we have used the fact that
\be
\hat D^1 \cdot \hat D^1 = \hat D^p \cdot \hat D^p
= {1 \ov p}-1, \qquad
\hat D^1 \cdot \hat D^p = -{1 \ov p} \,.
\ee
All the terms involving the strongly
coupled gauge fields are cancelled, as desired.
Observe that the coefficients of the $(\tr F^2)^2$
terms are quantized in units of $1/8p$.
When the flavor symmetries are gauged,
this coefficient contributes to the gauge anomaly
of the gauge groups $SU(N)$ and
$SU(N+pk)$.
Recall that the $(\tr F^2)^2$ terms of vector
or hypermultiplets are quantized in units of
$1/8$ \cite{KMT2}. Hence the contribution of the
superconformal matter to the gauge anomaly
can be interpreted as contributing in fractional
units of $p^{-1}$.
In the following section, we present examples where
these SCFTs appear in global F-theory backgrounds.

\section{F-theory on $X \ra \bP^2 / \bZ_3$}
\label{s:p2z3}

In this section, we study a particular example
of an F-theory compactification on a manifold
whose base has orbifold singularities. We consider an elliptically
fibered Calabi-Yau threefold $X \ra B$, where
$B = \bP^2/\bZ_3$. The orbifold group action is
defined such that it acts on the
projective coordinates $(U,V,W)$ of the
$\mathbb{P}^2$, which we often denote $\tB$, by
\be
(U,V,W) \ra (\om U, \om^2 V, W)
\label{oaction}
\ee
for the cube-root of unity $\om$. This orbifold
has three fixed-points at $U=V=0$, $V=W=0$
and $W=U=0$, where the geometry is locally
a
\be
\mathbb{C}^2 / \mathbb{Z}_3 \times T^2 \,.
\ee
At each fixed point, the action involves both primitive third roots
of unity, so this theory has three $(2,0)$ $A_2$
theories sitting at these loci.\footnote{This is because the action could
also be written in the two equivalent forms $(U,V,W)\ra (U, \om V, \om^2 W)$
and $(U,V,W) \ra (\om^2 U, V, \om W)$.}

This orbifold of $\bP^2$ is useful
to think about for a number of reasons.
Its cohomology is simple to describe,
and all the orbifold points are codimension-two.
Furthermore, when the complex structure of the
manifold is generic, the discriminant locus avoids
all the orbifold singularities. Also, there are points in
the complex structure moduli space where the gauge
seven-brane does intersect these orbifold points.
Hence, it is useful to investigate how
certain superconformal sectors show up as we tune
the complex structure moduli; for example,
we can move a seven-brane on top of the orbifold fixed point.
The methods for analyzing this particular model,
however, are expected to generalize to other bases
that have more complicated SCFTs generically.
%We comment more on directions to extend
%our results in the concluding remarks.

The complex structure deformations of this
model are represented by the allowed coefficients
of the usual Weierstrass model of the elliptic fibration
over $\mathbb{P}^2$ that are invariant under the
orbifold action \eq{oaction}. Recall that the
Weierstrass coefficients are given, in projective
coordinates, by
\be
f_{12} = \sum_{l+m+n=12} f_{l,m,n} U^l V^m W^n,\quad
g_{18} = \sum_{l+m+n=18} g_{l,m,n} U^l V^m W^n \,.
\label{W-coeff}
\ee
In order for $f_{12}$ and $g_{18}$ to be invariant
under the $\bZ_3$ action, only coefficients
of terms with
\be
l \equiv m \equiv n \mod{3}
\ee
are allowed to be nonzero. %It is simple to verify the number of such coefficients is given by $95$.
Thus, there are $95$ nonzero coefficients in \eq{W-coeff}.
To get the number of complex structure deformations
of the orbifold, we must subtract the number of
automorphisms of the fibration, which is given
by a $(\bC^*)^3$ action.
Two of the automorphisms are those of the base
that leave the fixed point invariant---they act on
$U/W$ and $V/W$. The other is an overall scaling
of the base.
The $(\bC^*)^3$ action then can be understood as
a rescaling of the three homogeneous coordinates.
We hence arrive
at $92$ complex structure deformations, all of which
can be identified with hypermultiplets. Then, the total
number of hypermultiplets in the theory becomes
$93$, by adding the hypermultiplet controlling the size
of the base manifold.
Meanwhile, the homology $H_2 (B,\bR)$ of the base manifold
is generated by a single element, that can be lifted to
three times of the
hyperplane class $3H$ of the covering $\bP^2$.
Therefore, the effective theory of this compactification
on a generic point in complex moduli space is given by
a $(1,0)$ supergravity theory with no tensor multiplets,
$93$ hypermultiplets and three $A_2$ theories.

Let us confirm that the anomalies are cancelled in
this effective theory.
The computation of
\cite{FHMM,HMM} implies that
for $Q$ coincident M5-branes,
\be
I_{Q} ={Q \ov 48} (-p_2 + {1 \ov 4} p_1^2) \,.
\ee
The anomaly of the $A_2$ theory can be obtained
from $I_{Q=3}$ by subtracting the anomaly of
a free (2,0) tensor multiplet---it is given by
\be
I_{A_2} = I_{Q=3} - I_{Q=1} = {1 \ov 24} (-p_2 + {1 \ov 4} p_1^2)
= {1 \ov 96} (\tr R^4 - {1 \ov 4} (\tr R^2)^2)
\ee
%Substituting
%\be
%p_1 =-{1 \ov 2} \tr R^2, \quad
%p_2 = -{1 \ov 4} \tr R^4 +{1 \ov 8} (\tr R^2)^2 \,,
%\ee
%we arrive at
%\be
%I_{A_2} =  \,,
%\ee
which is precisely the anomaly polynomial of two
$(2,0)$ tensor multiplets.
Then, the total anomaly polynomial of the theory is
given by
\bea
I_\text{tot} &=
{1 \ov 5760} (93-273)
(\tr R^4  + {5 \ov 4} (\tr R^2)^2)
+{9 \ov 128} (\tr R^2)^2 + 3 I_{A_2}
= {3 \ov 128 } (\tr R^2 )^2 \,.
\eea
This anomaly polynomial exactly coincides with
\be
{1 \ov 32 } \left({K \ov 2} \cdot {K \ov 2} \right)(\tr R^2 )^2
= {1 \ov 128 } \left({\widetilde{K} \cdot \widetilde{K} \ov 3}\right)
(\tr R^2 )^2  \,,
\ee
where we use tilded variables to denote divisors in the
manifold before orbifolding (which in this case is
$\mathbb{P}^2$), while untilded variables are used
to denote divisors in the orbifolds.
In this equation, $K$ is the canonical class of the base
orbifold while $\widetilde{K}$ is the canonical
class of $\bP^2$.
Recall that $\tK = -3H$, so that
\be
\tK \cdot \tK =9 \,.
\ee
Now given that the divisor $C_i$ of $B$
can be lifted to a divisor $\widetilde C_i$
in its covering space $\tB$,
the relation
\be
C_i \cdot C_j = {1 \ov 3} \widetilde C_i \cdot
\widetilde C_j
\label{1ov3}
\ee
holds. Thus we see that the anomalies of the theory
are cancelled by the GSSW anomaly
cancellation mechanism, with the gravitational anomaly
coefficient given by the canonical class of the base,
just as we have explained in the previous section.
%Note that the gravitational anomaly coefficient
%still has an interpretation as the canonical class of the
%base manifold, which in this case is an orbifold.
Note that the formula \eq{1ov3} suggests
that certain divisors in the base have fractional intersection
numbers. We explore the case when there is enhanced
gauge symmetry over such loci shortly.

Upon moving to special points in the complex
structure moduli space of $X$, the theory can acquire
enhanced gauge symmetry. For the sake of concreteness,
we only concern ourselves with $SU(N)$ gauge
symmetry---generalizations of our results
to other gauge groups is expected to be
straightforward. The effective
supergravity theory with $SU(N)$
gauge symmetry differs
depending on the nature of the
seven-brane locus $\s$ that carries the
gauge symmetry. In classifying the behavior of
$\s$, it is useful to examine the behavior of the
divisor $\ts$ on $\tB$ obtained by lifting
$\s$ to the cover of $B$. The irreducible
divisor $\s$ can then be one of the following:
\begin{enumerate}
\item The divisor $\s$ does not intersect any orbifold points.
  \begin{enumerate}
  \item $\ts$ is also irreducible on $\tB$.
  \item $\ts$ is reducible on $\tB$.
  \end{enumerate}
\item The divisor $\s$ intersects an orbifold point.
  \begin{enumerate}
  \item $\s$ is a Cartier divisor.
  \item $\s$ is Weil, but not Cartier.
  \end{enumerate}
\een

When $\s$ does not intersect any of the orbifold
points, the supergravity theory develops an $SU(N)$
gauge symmetry whose charged matter only
consist of hypermultiplets. As we show shortly,
when $\ts$ is irreducible on $\tB$, its genus is at
least one. Consequently, the genus of $\s$ is
also at least one, and the theory has an
adjoint hypermultiplet.
An interesting phenomenon
happens when $\ts$ is not irreducible
on $\tB$. In this case, $\ts$ factors into three
divisors, which project down to the single irreducible
divisor $\s$ on $B$. In this case, $\s$ develops
double points. Then, one of the global adjoint
hypermultiplets becomes localized at this point along
with a neutral hypermultiplet.

While a divisor $\s$ that does not intersect
any orbifold points is always Cartier, in the event
that $\s$ intersects an orbifold point, its defining equation
might not be a well-defined element of the
ring of rational functions
on the manifold. In the case that $\s$ is a Cartier divisor,
the point in complex
structure moduli space where $\s$ hits the orbifold point
can be approached from case 1-(a) by tuning the
complex structure modulus that controls the location
of $\s$.
As $\s$ hits the orbifold point, the $A_2$
theory is enhanced to the SCFT $\thyN$ %$ = T(SU(N),2)$
 with $SU(N) \times SU(N)$
global symmetry, whose diagonal group
is gauged by the $SU(N)$ gauge group. In fact, the
$A_2$ SCFT, an adjoint and a neutral
hypermultiplet are traded
for this new SCFT. It is interesting to understand
the string charge of the $SU(N)$ instantons of the theory.
These instantons are charged under the tensor degrees
of freedom in the SCFT, and hence the string charge lies
in the homology lattice of $\hat B$.
When $\s$ is not Cartier, something more drastic happens.
%With the notation of \cite{DHTV}, models of the type $T(SU(N),2,\mu_L,\mu_R)$ are produced.
The gauge divisor now has a fractional self-intersection
number---these fractional anomaly coefficients cancel the
anomalies of the SCFTs sitting at the orbifold loci,
which come in fractional units.

In the following, we examine each case in more detail.
Before doing so, however, we first describe the geometry
and topology of the manifold $B$ and its resolution $\hat B$
in more detail in subsection \ref{ss:B}.
In subsection \ref{ss:noSCFT}, we investigate the case when
the gauge divisor does not cross the orbifold locus. In subsection
\ref{ss:SCFT}, we discuss the case when it does.

\subsection{The geometry and topology of $\bP^2/\bZ_3$} \label{ss:B}

Let us review the geometry of the
orbifold $B$. The
{integral sublattice of the homology lattice}
of the orbifold is spanned by a single element $h$,
which lifts to three times the homology class of the
hyperplane in $\bP^2$:
\be
\widetilde h = 3H \,.
\ee
The self-intersection number of $h$ is given by
\be
h \cdot h = {1 \ov 3} (3H \cdot 3H) = 3\,.
\ee
Thus, the {integral sublattice of the homology lattice}
of $B$ is not unimodular.
$B$ has fractional divisors whose homology class
come in fractions of $h$. In fact, the basis vector
for the full homology lattice is given by $h/3$.

We focus our attention on
the Weil divisors $D_U$, $D_V$ and $D_W$
that come from projections of the divisors
$U$, $V$ and $W$ of $\tB$.
The homology class of these divisors are given by
\be
[D_U] = [D_V] = [D_W] = {1 \ov 3} h \,.
\label{homclass}
\ee
We use the square brackets to denote
the homology class of a divisor.
It is simple to see that
\be
D_x \cdot D_y = {1 \ov 3} (H \cdot H) = {1 \ov 3}
\ee
for any pair of $x, y \in \{ U, V, W\}$,
which is consistent with \eq{homclass}.

\begin{figure}[!t]
\centering\includegraphics[width=12cm]{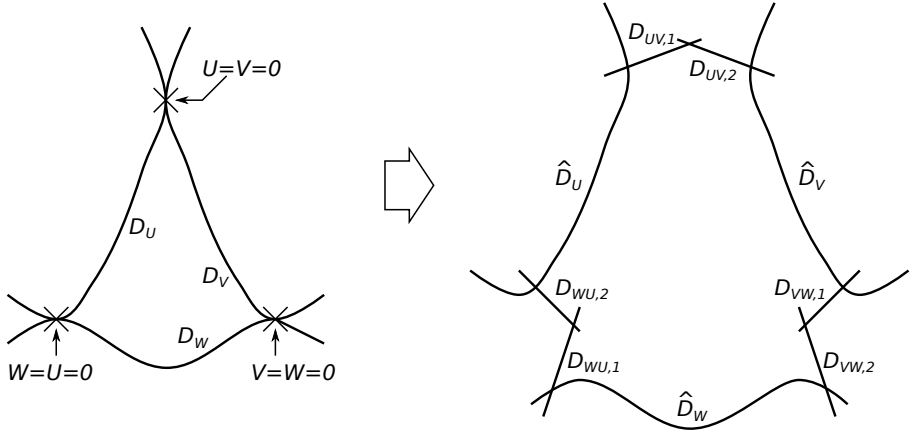}
\caption{\small $B = \bP^2 / \bZ_3$
and its resolution $\hat B$. $\hat B$ is a
$dP_6$.
There are six resolution divisors $D_{xy,a}$ that
resolve the three $A_2$ singularities of $B$.
The divisors $D_x$ of $B$ are mapped to
divisors $\hat D_x$. Each pair of adjacent
divisors in the diagram have intersection
number $1$.}
\label{f:UVWRes}
\end{figure}

Upon resolving the geometry by blowing up
the three $A_2$ singularities of $B$, we arrive
at $\hat B$, which is a del Pezzo surface of
degree three, or equivalently, a $dP_6$ manifold.
This blow up can be interpreted as going on
the tensor branch of the three SCFTs
localized at the three orbifold points.
There are different resolutions of the orbifold
singularities that are related to each other by flops.
These
flops correspond to going to different chambers
of the tensor branch of the SCFT.
To be unambiguous, we choose a particular
resolution in the succeeding discussions,
but it is straightforward to incorporate flops into
the picture. Upon resolving the singularities,
the divisors $D_x$ map into
divisors $\hat D_x$ of the manifold $\hat B$.
Let us denote the two resolution divisors that come
from resolving the singularity at $x=y=0$
by $D_{xy,a}$ with $a=1,2$.
The tensor branch parameters, in this particular
chamber, can be
identified with the sizes of the six cycles
$D_{xy,a}$. This resolution is depicted in
figure \ref{f:UVWRes}.

The seven homology classes
\be
[\hat D_U],~ [D_{UV,1}], ~[D_{UV,2}], ~\cdots~,
~[D_{WU,2}]
\label{hatBbasis}
\ee
form a basis for $H_2 (\hat B, \bR)$, though
it does not quite span the full integral homology lattice
of $\hat B$.
$\hat D_x$ are rational curves with self-intersection $(-1)$,
while the resolution divisors are rational curves with
self-intersection $(-2)$.
Hence the intersection matrix of the divisors \eq{hatBbasis}
can be read off of the diagram on the right-hand-side
of figure \ref{f:UVWRes}.
A more relevant basis for $H_2 (\hat B, \bR)$
for our discussion, of course, is one that is spanned by
$D_{xy,a}$ and an integral homology class orthogonal
to $D_{xy,a}$, namely
\be
h = 3 [\hat D_U]
+ {2 } [D_{WU,2}] +  [D_{WU,1}]
+{2 } [D_{UV,1}] + [D_{UV,2}] \,.
\label{h in hatB}
\ee
This class has self-intersection $3$, while
it is orthogonal to all the resolution divisors,
and hence can be identified with
the basis of the
{integral sublattice of the homology lattice}
of $B$.
Since all the divisors $D_{xy,a}$ are $(-2)$
rational curves,
the canonical class $\hat K$ of $\hat B$
is orthogonal to these---in fact,
its homology class is given by
\be
[\hat K ] = -h \,.
\ee

The integral homology lattice $H_2 (\hat B, \bZ)$
is spanned by seven elements, $e_i$
with $i= 0, \cdots,6$.
A $dP_6$ surface can be thought of as a
smooth $\bP^2$ blown up at six generic points.
The element $e_0$ is the hyperplane class of the
original $\bP^2$, while the six elements $e_i$
are the exceptional cycles coming from the blow-ups.
The intersection matrix between these elements
is given by
\be
(e_i \cdot e_j) = \text{diag} \,(1, -1, -1, -1,-1,-1,-1) \,.
\ee
The elements $h$ and $[D_{xy,a}]$ can be related
to $e_i$ by
\bea
h&=3 e_0 - e_1 - e_2 - e_3 - e_4 - e_5 - e_6, \\
[D_{UV,1}]&= -e_1 +e _2, \quad
[D_{UV,2}]=-e_2+e_3, \quad
[D_{VW,1}]=e_4-e_5, \quad
[D_{VW,2}]=e_5-e_6, \\
[D_{WU,1} ]&= -e_0 + e_1 + e_2 + e_3 , \quad
[D_{WU,2} ]= -e_0 + e_4 + e_5 + e_6 \,.
\eea

Now while the homology classes of $D_U$, $D_V$
and $D_W$ were the same in $B$, we observe that
the homology classes
of $\hat D_U$, $\hat D_V$ and $\hat D_W$ differ.
In fact,
\begin{align}
[ {\hat D_U} ] &= {1 \ov 3} h - {1 \ov 3} [D_{WU,1}]- {2 \ov 3} [D_{WU,2}]
-{2 \ov 3} [D_{UV,1}] -{1 \ov 3} [D_{UV,2}] \\
[ {\hat D_V} ] &= {1 \ov 3} h - {1 \ov 3} [D_{UV,1}]- {2 \ov 3} [D_{UV,2}]
-{2 \ov 3} [D_{VW,1}] -{1 \ov 3} [D_{VW,2}] \\
[ \hat D_W] &={1 \ov 3} h - {1 \ov 3} [D_{VW,1}] - {2 \ov 3} [D_{VW,2}]
-{2 \ov 3} [D_{WU,1}] -{1 \ov 3} [D_{WU,2}] \,.
\label{DUVW}
\end{align}
It is a simple exercise to check that the self-intersection
numbers of these homology classes are indeed given by $(-1)$.
The fractional coefficients reflect the fact
that $h$ and $D_{xy,a}$
do not span the full integral homology lattice of $\hat B$.
When these divisors are written as a linear combinations of
the basis elements $e_i$, in fact, the coefficients are
integral.
It is evident from this formula that the projection of
$\hat D_x$ down to the sublattice of $H_2 (\hat B,\bZ)$
spanned by $h$ all become $h/3$.
The homology class of Cartier divisors $D$
of $B$ that do not intersect
orbifold loci in the class $[D] = nh$ remain the
same through the blow up.
Meanwhile, as we see in section \ref{ss:SCFT},
Cartier divisors
that intersect an orbifold locus may contain
the resolution divisors as componenets upon
blowing up the orbifold points.

Let us explore the physical implications of the facts
presented. Given a generic elliptic fibration $X \ra B$,
there exists BPS strings of the six-dimensional F-theory
compactification on $X$ obtained by wrapping D3-branes
on divisors of $B$. If the D3-brane is wrapping a
divisor $D$ that intersects an orbifold locus,
it is not enough to know the homology class of
$D$ in $B$ to determine its full string charge. The full string
charge is given by the homology class of $\hat D$ in
$H_2 (\hat B, \bZ)$. While the homology class of the
three divisors $D_{U,V,W}$ in $B$ are equivalent,
we see explicitly from equation \eq{DUVW} that
the homology classes of $\hat D_{U,V,W}$ differ.

Meanwhile, as we have shown in the previous section,
given that there is a gauge brane wrapping a divisor of $B$,
the anomaly coefficient of the gauge group still can be
identified with the homology class of that divisor within $B$.
For example, the anomaly coefficient of a gauge group
obtained by wrapping a brane on $D_U$ is given by $h/3$.
The string charge of a unit instanton of that gauge
group, however, is given by the homology class of $\hat D_U$ in \eq{DUVW}.
As noted before, when a gauge brane is wrapping a
divisor intersecting an orbifold point,
the anomaly coefficient cannot be identified with
the string charge of the unit instanton of that gauge group.

\subsection{Enhanced gauge symmetry without charged
superconformal matter}
\label{ss:noSCFT}

Let us now consider loci of the complex moduli space of $X$
where there is an $SU(N)$ gauge symmetry along
the locus $\s$. The lift of $\ts$ of $\s$ to $\tB = \bP^2$
must be of the form:
\be
\ts = p( u^3 , v^3 , uv ) \,,
\ee
where $p$ is a polynomial in three variables, and
$u$ and $v$ are local coordinates of the
$\bP^2$ in the chart $W=1$. The class of $\ts$
hence is always given by a multiple of $3H$, consistent
with the fact that then
\be
\s \cdot \s = {\ts \cdot \ts} / 3, \quad
\s \cdot K = {\ts \cdot \tK} / 3
\label{intov3}
\ee
is integral. The genus $g$ of $\s$ is
then given by
\be
g = {1 \ov 2}(\s \cdot \s + K \cdot \s +2) =
{1 \ov 6}(\ts \cdot \ts-3H \cdot \ts+6) \,.
\label{genus}
\ee

The gauge theory on $\s$ can be
thought of as a ``quotient" of the gauge theory
on $\ts$ living on $\tB$, in the following sense.
The gauge theory on $\ts$ has
\be
\tg = {1 \ov 2}(\ts \cdot \ts-3H \cdot \ts+2) = 3g -2
\ee
global adjoint hypermultiplets \cite{WittenPT,KatzMorrisonPlesser}.
The other matter come from loci where $\ts$
intersects the rest of the discriminant locus, or
where $\ts$ itself develops a singularity.
Now the Weierstrass equation of the theory on
$\tB$ must be restricted to be invariant under the
$\bZ_3$ action.
Then, since the $\bZ_3$ action acts freely on the locus
$\ts$, as it does not cross through any orbifold points,
such loci come in triplets. Hence the
rest of the matter, other than the $\tg$ adjoint
matter, come in triplets.
Upon quotienting by $\bZ_3$,
the gauge theory on $\s$ has $g$
adjoint hypermultiplets. The rest of
the charged matter spectrum
can be obtained by quotienting the charged matter
of the theory on $\tB$ that come from
codimension-two singularities by three,
as each triplet of singular loci on $\tB$
reduces to a single codimension-two
locus on $B$.

To be more concrete, let us consider the
class of supergravity theories with
$SU(N)$ gauge symmetry on $B$
that can be obtained by quotienting
a theory on $\bP^2$
with the following matter content
in the non-abelian sector
\cite{MTMatter}:
\be
SU(N): \;\;\;\;\; \;\;\;\;\;
(72-9N) \times {\tiny\yng(1)} +
9 \times {\tiny\yng(1,1)}+
1 \times {\rm Adj} \,,
\;\;\;\;\; [\ts] = 3H \,,
\ee
for $N \leq 8$. As indicated,
the cohomology class of $\ts$ in this case,
is given by three-times the hyperplane class.
The divisor $\ts$ can be written as
\be
\ts: \quad a u^3 + b v^3 + c  +d uv = 0 \,.
\label{ts}
\ee
We use $\ts$ to both denote the divisor itself
as well at its defining equation.
The genera $\tg$ and $g$
are given by
\be
\tg = g =1 \,.
\ee
Hence, the theory on $\bP^2$ has a single adjoint
global multiplet, while the rest of the matter
come from codimension-two singularities
that can be organized into triplets by acting
on the $\bP^2$ with the $\bZ_3$ action.
The non-abelian gauge group of the
theory on $B$ then, is given by $SU(N)$ under which
the representations of the charged matter content
are given by
\be
SU(N): \;\;\;\;\; \;\;\;\;\;
(24-3N) \times {\tiny\yng(1)} +
3 \times {\tiny\yng(1,1)}+
1 \times {\rm Adj} \,,
\;\;\;\;\; [\s] = h \,.
\label{suN family}
\ee
Recall that $h$ is the homology
class on $B$ that lifts to the homology class of
$3H$ on $\tB$, i.e.,
\be
\widetilde h =3 H.
\label{defh}
\ee
Note that
\be
h \cdot h = 3, \quad h \cdot K = - 3\,.
\ee
The theory has
three $A_2$ SCFTs coupled to the
gravity theory as well.
The gauge and mixed anomaly cancellation conditions
\bea
{1 \ov 16} (K \cdot h) \tr R^2 \tr F^2
&=
-{1 \ov 96} \tr R^2 \left[ (24-3N) \tr F^2 + 3 \tr_{\tiny\yng(1,1)} F^2 +
 \tr_\text{Adj} F^2 -  \tr_\text{Adj} F^2 \right]  \\
{1 \ov 8} (h \cdot h) (\tr F^2)^2
&={1 \ov 24}
 \left[  (24-3N) \tr F^4 + 3 \tr_{\tiny\yng(1,1)} F^4 +
 \tr_\text{Adj} F^4 -  \tr_\text{Adj} F^4\right]
\label{suN GSSW}
\eea
involving the $SU(N)$ gauge group is
then satisfied.
The relations
\be
\tr_{\tiny\yng(1,1)} F^2 = (N-2) \tr F^2, \quad
\tr_{\tiny\yng(1,1)} F^4 = (N-8) \tr F^4 + 3 (\tr F^2)^2
\ee
are needed to show the equality \eq{suN GSSW}.
We also note that
\be
\tr_\text{Adj} F^2 = 2N \tr F^2, \quad
\tr_\text{Adj} F^4 = 2N \tr F^4 + 6 (\tr F^2)^2
\ee
for future reference.
The anomaly coefficient of the $SU(N)$ gauge
group can hence be identified as $h$, which is the
class of the $SU(N)$ divisor $\s$.

It is possible to verify the gauge and mixed
anomaly equations for any theory on $B$
whose gauge brane $\s$
is smooth, irreducible and does not cross the
orbifold locus.
This is because the lifted theory to $\tB$,
which is a theory with gauge group
$G$, $\tg$ global adjoint
hypermultiplets and $3 n_{\bf R}$ hypermultiplets in
the representation $\bf R$, is also a consistent theory
that must satisfy the gauge and mixed
anomaly equations
\bea
{1 \ov 16} (\tK \cdot \ts) \tr R^2 \tr F^2
&=
-{1 \ov 96} \tr R^2 \left[ \sum_{\bf R} 3n_{\bf R} \tr_{\bf R} F^2
+ (\tg -1)\tr_\text{Adj} F^2 \right]  \\
{1 \ov 8} (\ts \cdot \ts) (\tr F^2)^2
&={1 \ov 24}
 \left[  \sum_{\bf R} 3n_{\bf R} \tr_{\bf R} F^4 +
(\tg-1) \tr_\text{Adj} F^4\right] \,.
\eea
The gauge and mixed anomaly cancellation
equations for the
theory on $B$ can be obtained from these equations
by dividing both sides of the equations by three.
This follows from the fact that the charged matter
content of $B$ consists of
\be
g = {1 \ov 3}({\tg -1})+1
\ee
global adjoint hypermultiplets and $n_{\bf R}$
hypermultiplets that come from local singularities,
and that the intersection numbers between the
anomaly coefficients $\s$ and $K$ of the theory
are related to those of $\ts$ and $\tK$ by \eq{intov3}.
This proof generalizes straightforwardly to
any supergravity theory on $B$, with any gauge group
(with multiplet semi-simple factors) whose gauge divisors
are smooth, irreducible in $\tB$ and avoid all the
orbifold points.

The number of neutral hypermultiplets $\nu_H$ of the
theories \eq{suN family} can be computed using
the gravitational anomaly constraint:
\be
H-V +29 T + \Delta_{S} = 273 \,,
\ee
where $H$/$V$/$T$ are the number of
hyper/vector/tensor multiplets of the theory,
and $\Delta_S$ is the contribution of the strongly
coupled sector to the gravitational anomaly.
In the event that none of the gauge branes
cross the orbifold loci, we have computed
\be
\Delta_S = 3 \times 60 = 180 \,,
\ee
since each of the three $(2,0)$ $A_2$ theories
contribute to the gravitational anomaly
as much as two hyper and two tensor multiplets do.

Now the F-theory models \eq{suN family} can have
abelian factors, given that the abelian gauge group
cannot be Higgsed away without breaking the
non-abelian gauge group.
As we explain in more detail in appendix \ref{ap:coefficients},
it turns out some members of the family of F-theory models
\eq{suN family} \textit{automatically} have an additional
$U(1)$ factor, all of whose charged particles have
non-trivial non-abelian representations (see
table \ref{t:coefficients}).
Hence, for this class of theories, the number
of vector and hypermultiplets are given by
\be
V= r_{MW}+ N^2 -1 \,,
\quad
H =
\begin{cases}
\nu_H+36 \,, &\text{when $N=2$} \\
\nu_H-{1 \ov 2} N^2 +{45 \ov 2} N-1 \,, & \text{when $N \geq 3$} \,
\end{cases},
\ee
where $r_{MW}$ is the number of abelian factors,
or equivalently, the Mordell-Weil rank of the elliptic
fibration \cite{MV2}.
The counting of charged matter is slightly
different for the theory with an
$SU(2)$ gauge group, as the antisymmetric
representation is neutral in that case.
The gravitational anomaly constraint shows
that the number of neutral hypermultiplets is given by
\be
\nu_H =
\begin{cases}
57 \,, &\text{when $N=2$} \\
93+{3 \ov 2} N(N - 15 ) + r_{MW} \,, & \text{when $N \geq 3$} \,.
\end{cases}
\label{neutral hypers}
\ee

It is quite interesting to compare the
number of neutral hypermultiplets of the
$SU(N)$ theories with the number of free
complex coefficients of the Weierstrass model.
As before, the number of complex
coefficients can be enumerated
by counting the number of complex coefficients
of Weierstrass models over
the covering space $\bP^2$ that are $\bZ_3$-%
invariant. The general form of Weierstrass
models with enhanced $SU(N)$ symmetry has
been systematically studied in \cite{MTMatter}.

\begin{table}[t!]
\center
 \begin{tabular}{ | c || c | c | c | c| c| c| c| }
 \hline
 $N$ & 2 & 3 & 4 & 5 & 6 & 7& 8 \\ \hline\hline
 $w$ & 56  & 38 & 26 & 17 & 12 & 9 & 8 \\ \hline
 $\nu_H$ & 57  & 39 & 27 & 18 & 13 & 10 & 9 \\ \hline
 $r_{MW}$ & 0  & 0 & 0 & 0 & 1 & 1 & 0 \\ \hline
  \end{tabular}
 \caption{The number of free complex parameters
 of Weierstrass models $w$ vs. the number of neutral
 hypermultiplets $\nu_H$
 for F-theory models on $\bP^2/\bZ_3$
 whose non-abelian gauge group is given by
 $SU(N)$. The total gauge group of the theory
 is given by $SU(N) \times U(1)^{r_{MW}}$.
 For the $SU(6)$ and $SU(7)$
 cases, the abelian gauge group is non-trivial.}
\label{t:coefficients}
\end{table}

We conclude this section by enumerating
the number of free complex parameters for
the $SU(2)$ theory and comparing with the number
of neutral hypermultiplets \eq{neutral hypers}, following \cite{MTMatter}.
The details of the general counting for $SU(N)$
with $N \leq 8$ is presented in appendix \ref{ap:coefficients},
while the results are collected in table \ref{t:coefficients}.
 By the usual arguments \cite{MV1,KMT1,MTMatter},
the complex degrees of freedom of the Weierstrass
model can represent at most $(\nu_H-1)$ neutral
hypermultiplets. The additional hypermultiplet comes
from the overall scaling of the base.
We see that this bound is saturated
for all of the models we examine.
According to \cite{MTMatter},
Weierstrass coefficients of a manifold
that has an $SU(2)$
singularity along the locus $\ts$
must be of the form:
\bea
f&= -{1 \ov 48}\phi^2 +  f_1 \ts + f_2 \ts^2 \\
g&={1 \ov 864} \phi^3 -{1 \ov 12} \phi f_1 \ts + g_2 \ts^2 \,.
\label{su2coeff}
\eea
$\bZ_3$-invarance imposes that
all individual factors appearing in this expression
should also be $\bZ_3$-invariant, since $\ts$ is chosen
to be a $\bZ_3$-invariant locus by assumption.

Now we can fix the coefficients of $\phi$ and $f_1$
such that it is of the form
\bea
\phi & =  \varphi_0 u^2 v^2 +\varphi_{3} (v^3) uv + \varphi_6 (v^3) \\
f_1 &= f_{1,3} (v^3) u^2v^2 + f_{1,6} (v^3) uv + f_{1,9} (v^3) \,.
\eea
This is done by replacing any multiple of $u^3$
appearing in $\phi$ or $f_1$
by using the relation \eq{ts}.
The polynomials $\varphi_{3n}$, $f_{1,3n}$
are polynomials of $v^3$ of degree $3n$
in $v$.
For example, $\varphi_6 (v^3)$ is of the form
\be
\varphi_6 = p_6 v^6 + p_3 v^3 + p_0 \,,
\ee
and so on. Meanwhile, $f_2$ and $g_2$
are generic $\bZ_3$ invariant polynomials with
maximum degree 6 and 12 in the local coordinate
variables, respectively;
\be
f_2 (u^3,v^3,uv), \quad g_2 (u^3,v^3,uv) \,.
\ee
Summing all the number of
complex coefficients present in the model,
including the coefficients of $\ts$ \eq{ts},
we find $60$ complex coefficients in total.
Now there is a rescaling symmetry that leaves $f$
and $g$ invariant, given by
\be
\ts \ra t \ts, \quad
f_1 \ra t^{-1} f_1, \quad
f_2 \ra t^{-2} f_2, \quad
g_2 \ra t^{-2} g_2 \,.
\ee
This symmetry, along with the $(\bC^*)^3$
automorphism group of the elliptic fibration, cuts
down the number of free complex coordinates to $56$,
which agrees with $(\nu_H-1)$ of the theory.

The discriminant locus of the Weierstrass model
with the coefficients \eq{su2coeff} is of the form
\be
{1 \ov 16}\ts^2 \Big\{
\phi^2 \left({1 \ov 12} f_2 \phi^2 +g_2 \phi -f_1^2\right) + \cO (\ts)
\Big\} \,.
\ee
In $\tB=\bP^2$, the $A_1$ singularity on $\ts$ is enhanced
to $A_2$ at the $18 \times 3 =54$ points
where $\ts$ and
\be
{1 \ov 12} f_2 \phi^2 +g_2 \phi -f_1^2 = 0
\ee
intersect. These points come in triplets, which are
exchanged amongst themselves upon acting with
the $\bZ_3$ action. Hence, in $B$, there are
$18$ points where the $I_2$ fiber along $\s$
enhances to an $I_3$ fiber. These points
are where the $18$
fundamental matter of the $SU(2)$ group are
localized on. There is no additional matter
lying at the loci where $\ts$ and $\phi$ meet,
as the $I_2$ fiber along $\ts$
becomes a type $III$ fiber at these loci---there
is no increase of rank in this case.

\begin{figure}[!t]
\centering\includegraphics[width=9cm]{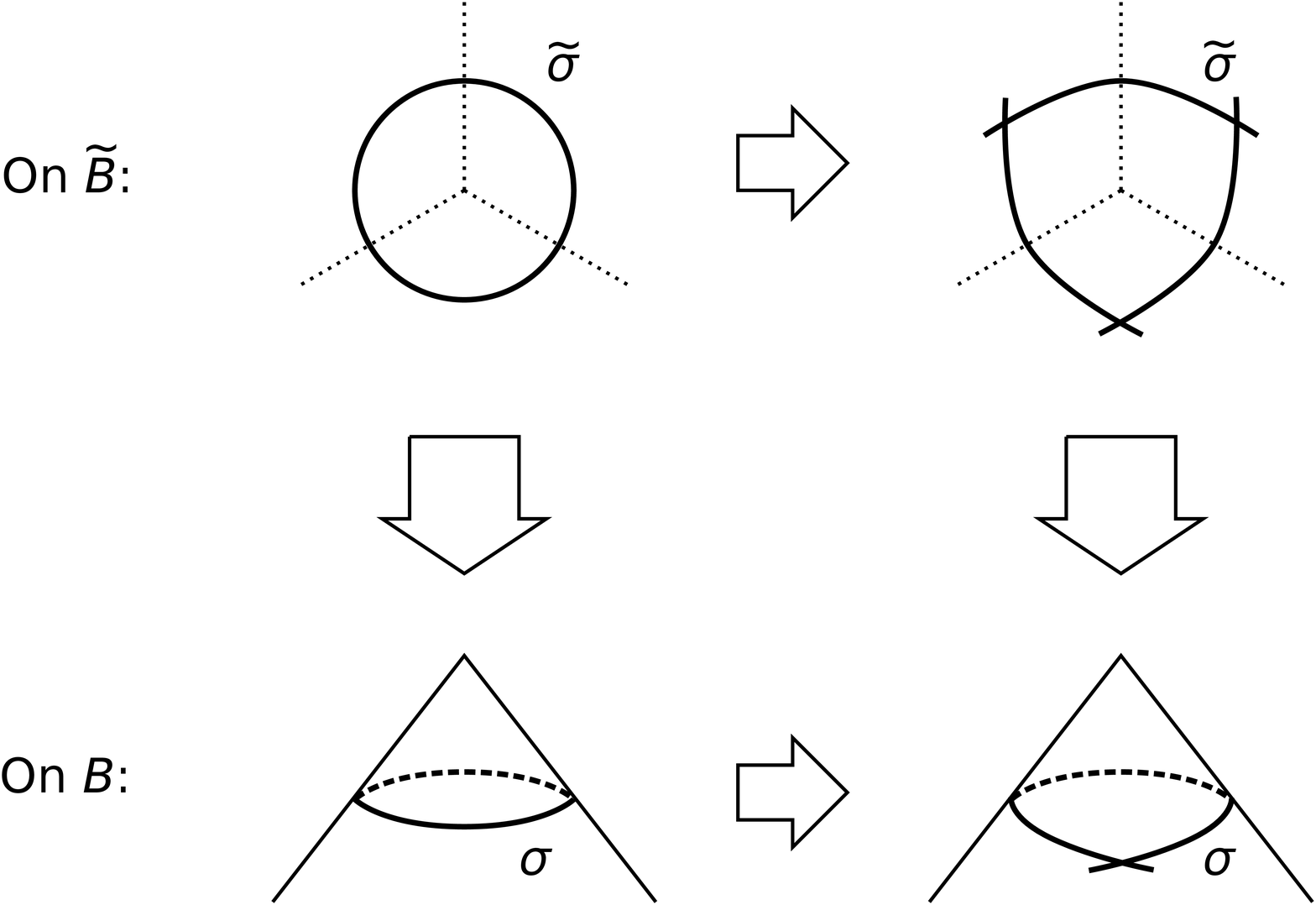}
\caption{\small A schematic picture of $\ts$
as it becomes reducible in $\tB$. The upper
diagrams depict the locus of the
divisor $\ts$ (bold curves) on $\tB$, while
the lower diagrams depict its projection,
$\s$ (also bold curves) on $B$.
The orbifold $B$ is depicted as a cone, while
the dotted lines on $\tB$ are used to
show the fundamental domain of $\tB$
under the orbifold action.
When $\ts$ is irreducible (left),
its projection is a smooth divisor on $B$.
Meanwhile, when $\ts$ become reducible (right),
it factors into three copies of divisors related by
the $\bZ_3$ action. Upon projection to $B$,
$\s$ develops a double-point.}
\label{f:factorization}
\end{figure}

When $\ts$ becomes factorizable, i.e.,
when the coefficients of its defining equation
\eq{ts} satsify
\be
d^3 + 27 abc =0 \,,
\label{modulus}
\ee
an interesting situation occurs.
The factors of $\ts$,
\be
\ts =  (a^{1/3} u + b^{1/3} v + c)
(\om a^{1/3} u + \om^2 b^{1/3} v + c)
(\om^2 a^{1/3} u + \om b^{1/3} v + c) \,,
\ee
are not invariant divisors on $B$. Therefore the
divisor $\s$ is still irreducible on $B$.
Meanwhile, each pair of
the three factors of $\ts$ meet with
each other at a point. The three intersection points
of these divisors on $\tB$
project to a single double point of $\s$ in $B$.
The situation is sketched in figure \ref{f:factorization}.
In principle, when a gauge brane is wrapping a curve
with a double point,
the matter localized at the double point locus
can either be given by a pair of hypermultiplets
in the adjoint and the trivial representations,
or in the symmetric and anti-symmetric representations
\cite{MTMatter}.
In this case, however, the double point locus
can be reached by a continuous deformation
of the parameters in $\s$. We can therefore conclude
that the matter localized at the double point is
an adjoint and a neutral hypermultiplet.
The neutral hypermultiplet can be identified with
the combination \eq{modulus} of coefficients of $\ts$.
The effective theory on $B$ hence remains the same
at this locus, despite the development of the singularity
on $\s$.

\subsection{Enhanced gauge symmetry with charged
superconformal matter}
\label{ss:SCFT}

We now examine the case when the $SU(N)$
gauge brane passes through an orbifold locus.
Let us first consider the case that
the gauge brane is wrapping a
Cartier divisor $\s$.
This situation arises by
starting at a point in the complex
structure of the moduli space of $X$
where the gauge brane locus $\s(c)$
is a Cartier divisor
that does not intersect the orbifold
point. Here we have used $c$ to denote
the parameter of $\s(c)$
that needs to be tuned to reach the orbifold
point. Then, we can make the gauge brane
cross the orbifold singularity by tuning
the coefficient $c$ to a particular value $c_0$.
For example, for the class of $SU(N)$ theories
studied in the previous subsection,
we can tune the coefficient $c$ of the equation
\eq{ts} to zero so that
\be
\ts(c=0) : \quad au^3 + bv^3 + d uv =0
\ee
passes through the orbifold singularity
at $u=v=0$.
In fact, given that we want to make the divisor
hit this orbifold point, there is a single coefficient
$c$ that we need to tune to zero to do so---the constant
term in $\ts(c)$, when $\ts$ is written as a polynomial
in $u$ and $v$.

Let us assume that the supergravity theory
with non-zero $c$ had $g$ global adjoint
hypermultiplets and $n_{\bf R}$ local hypermultiplets
for each representation ${\bf R}$ of the gauge group
$SU(N)$. Upon tuning $c$ to zero, a global
adjoint hypermultiplet, the $A_2$ theory
sitting at the orbifold point, along with the neutral
hypermultiplet degree of freedom parametrized by $c$,
enhances into a $\thyN$ SCFT whose diagonal
$SU(N)$ global symmetry group is gauged. The rest of the matter
remains the same, as the local codimension-two
singularities merely shift their positions as $c$
is taken to zero.
%%%%Figure of the situation%%%%

Now let us check the anomaly cancellation conditions
are satisfied for this theory.
Assuming that the anomaly of the theory with $c \neq 0$
is cancelled, it follows that the anomaly of
the theory with $\thyN$ is also cancelled with
the same anomaly coefficients, due to the relation
\bea
I_S &= {1 \ov 5760} (60+N^2) \tr R^4 -{2 \ov 128} (\tr R^2)^2 \\
&-{1 \ov 96} (2N) \tr R^2 \tr F^2+{1 \ov 24} (2N \tr F^4+6(\tr F^2)^2) =
I_{A_2}+I_{H,\text{Adj}}+I_{H,\text{neutral}} \,,
\eea
where $I_S$ is the anomaly polynomial of the SCFT
$\thyN$, computed in section \ref{fracanoX}. The anomaly polynomials for the traded
fields add up precisely to the anomaly polynomial of the
SCFT!
The anomaly polynomial $I_S$ can be obtained
from equation \eq{anom Nkp} by setting
\be
F_0 = F_p =F \,,
\ee
for the gauge fields strength $F$ of the $SU(N)$ group.

\begin{figure}[!t]
\centering\includegraphics[width=10cm]{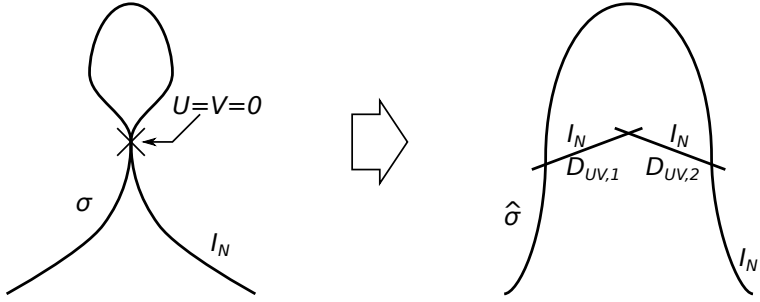}
\caption{\small A diagram depicting the resolution of
the $A_2$ singularity at $U=V=0$ when the gauge brane
$\s$ carrying an $I_N$ singularity passes through.
The two resolution divisors
$D_{UV,1}$ and $D_{UV,2}$ each have a $I_N$
singularity along them.}
\label{f:T3NN}
\end{figure}

Let us now move to a generic point in
the tensor branch of the SCFT $\thyN$.
By doing so, we resolve the $A_2$ singularity
at $U=V=0$, as shown in figure \ref{f:T3NN}.
As noted previously, when a Cartier
divisor of $B$ does not intersect an orbifold locus,
its homology class remains in the sublattice of
$\hat B$ spanned by $h$ even after the blow up.
In the case we are considering, however, $\s$
intersects the orbifold locus---the divisor $\hat \s$
obtained by blowing up $\s$ contains $D_{UV,a}$
as its components.
This is signified by the fact that $\hat \s$ has non-zero
intersection numbers with the resolution divisors,
as can be seen in figure \ref{f:T3NN}. The homology
class of $\hat \s$ is given by
\be
[\hat \s] = h - [D_{UV,1}]- [D_{UV,2}] \,.
\ee
Note that
\be
\hat \s \cdot  \hat \s = 1, \quad
\hat \s \cdot \hat K = -3 \,,
\label{Cartier hat s}
\ee
The charged matter with respect to the
$SU(N)$ gauge symmetry living on
$\hat \s$ is given by the $(24-3N)$ fundamentals
and the three antisymmetrics along with
two bifundamentals each living at the intersection
point between $\hat \s$ and $D_{UV,1}$, $D_{UV,2}$.
The gauge and mixed anomaly cancellation
conditions for this gauge group component is
consistent with the intersection numbers \eq{Cartier hat s}.

By inspection of the geometry, we see that while the anomaly
coefficient of the $SU(N)$ gauge group on $B$
should be identified with $h \in H_2 (B, \mathbb{Z})$, the string
charge of the instantons of the gauge group cannot
lie within $H_2 (B, \mathbb{Z})$. The instantons are
charged under tensor multiplets
of the SCFT $\thyN$, as we see that the divisor $\hat \s$
intersects the resolution divisors $D_{UV,a}$.
This should be contrasted to the case when a Cartier
divisor of $B$ does not intersect an orbifold locus, in which
the divisor does not contain any resolution divisors as a
component.
The string charge of the unit $SU(N)$ instanton
should be identified
with the homology cycle
of $\hat \s$ given by the element
\eq{Cartier hat s}
of the homology lattice of $\hat B$.

Let us now discuss what happens
when the gauge brane
wraps a Weil divisor $\s$ that is not
Cartier. In this case, the intersection numbers of the
brane become fractional. For the remainder of the
section, we explore the case where the Calabi-Yau
manifold has $I$-type
singularities along the simplest fractional
divisors---those that can be lifted to $U=0$, $V=0$ and
$W=0$. As before, let us denote these divisors on $B$,
$D_U$, $D_V$ and $D_W$, respectively.
Recall that these
divisors have fractional intersection numbers:
\be
D_x \cdot D_y = {1 \ov 3} , \quad
D_x \cdot K = -1
\label{Dxint}
\ee
for any pair of $x, y \in \{ U, V, W\}$.
The homology class of $D_x$ are all given by $h/3$.
Upon lifting to the covering manifold $\bP^2$,
we obtain models with $I$-type singularities
along divisors in the hyperplane class $H$.
Such $SU(N)$ models
have been studied in \cite{MTMatter,KPT}. A particular
model that is simple to engineer is one with the
following matter:
\be
SU(N): \;\;\;\;\; \;\;\;\;\;
(24-N) \times {\tiny\yng(1)} +
3 \times {\tiny\yng(1,1)} \,.
\ee
Note that there is no adjoint hypermultiplet, as the seven-brane wraps a $\mathbb{P}^1$.
For example, such a theory can be engineered on $\bP^2$ with an $I_N$ singularity fibered
over $U=0$.
As explained in the previous section,
a brane configuration on $B$ cannot be lifted to
a configuration on $\tB$ with only an $I_N$ singularity
at $U=0$. In fact, given that there is an $I_N$ fiber
along $U=0$, there must be corresponding singular fibers at the other loci $V = 0$ and $W = 0$.
Owing to the $\mathbb{Z}_3$ invariance of the elliptic fibration, the total seven-brane charges must be the same mod 3, so we
can introduce integers $k$ and $k'$, and assign an $I_{N+3k}$ fiber
along $V=0$ and an $I_{N+3k'}$ fiber along
$W=0$. The discriminant locus $\Delta$ of this model on
$\bP^2$ becomes
\bea
\Delta &= U^N V^{N+3k} W^{N+3k'} \cF \,,
\eea
in projective coordinates, where $\cF$ can be written as
\bea
\cF & = \phi_{0,U}^4 \Phi_U  + \cO(U)
=  \phi_{0,V}^4 \Phi_V  + \cO(V)
=  \phi_{0,V}^4 \Phi_W  + \cO(W)  \,.
\eea
Here, $\phi_{0,x}$ are sections
of $3H$, while $\Phi_x$ are sections of $3(8-N-k-k')H$.
F-theory compactified on this Calabi-Yau fibration over $\bP^2$
would yield a supergravity theory with gauge group
\be
G_U \times G_V \times G_W \equiv SU(N) \times SU(N+3k) \times SU(N+3k')
\ee
and the following matter:
\bea
&(24-3N-3k-3k') \times \Big\{
({\tiny\yng(1)},\cdot,\cdot) +
(\cdot,{\tiny\yng(1)},\cdot) +
(\cdot,\cdot,{\tiny\yng(1)}) \Big\} \\
&+
3 \times \Big\{
({\tiny\yng(1,1)},\cdot,\cdot) +
(\cdot,{\tiny\yng(1,1)},\cdot) +
(\cdot,\cdot,{\tiny\yng(1,1)}) \Big\}
+ ({\tiny\yng(1)},{\tiny\yng(1)},\cdot)
+(\cdot,{\tiny\yng(1)},{\tiny\yng(1)})
+({\tiny\yng(1)},\cdot,{\tiny\yng(1)})
\,.
\eea
The fundamental matter come from the intersections
of gauge branes $D_x$ and $\Phi_x$,
while the antisymmetrics come from the intersections
between $D_x$ and $\phi_{0,x}$.
The bifundamental matter lie at the intersections
between gauge branes. As expected, the values of $k$ and $k^{\prime}$ are bounded above and below. For example, we have the upper bound:
\begin{equation}
3N + 3k + 3k' \leq 24
\end{equation}
and the lower bounds:
\begin{equation}
N \geq 0 \,\,\, N + 3k \geq 0 \,\,\, N + 3k' \geq 0.
\end{equation}

Upon orbifolding this theory, we arrive at
a theory on $B$ with an $I_N$, $I_{N+3k}$ and
$I_{N+3k'}$ singularity along $D_U$, $D_V$ and
$D_W$, respectively. The divisors $\phi_0$ and
$F$ become sections of $h$ and $(8-N-k-k')h$.
Each fractional divisor meets the projection of
$\phi_0$ at a point where an antifundmental
hypermultiplet lies, and the projection of
$F$ at $(8-N-k-k')$ points, where fundamental
multiplets are localized. The charged hypermultiplet
spectrum is thus given by
\bea
(8-N-k-k') \times \Big\{
({\tiny\yng(1)},\cdot,\cdot) +
(\cdot,{\tiny\yng(1)},\cdot) +
(\cdot,\cdot,{\tiny\yng(1)}) \Big\}+
\Big\{
({\tiny\yng(1,1)},\cdot,\cdot) +
(\cdot,{\tiny\yng(1,1)},\cdot) +
(\cdot,\cdot,{\tiny\yng(1,1)}) \Big\}
\,.
\eea
Meanwhile, at the orbifold points $U=V=0$,
$V=W=0$ and $W=U=0$, lies the strongly
coupled SCFTs $\thy{N}{k}{3}$,
$\thyNkkp$, and $\thy{N}{k'}{3}$,
whose global symmetries have now been weakly gauged.
A schematic diagram of the configuration
of the divisors is given in figure \ref{f:UVW}.

\begin{figure}[!t]
\centering\includegraphics[width=14cm]{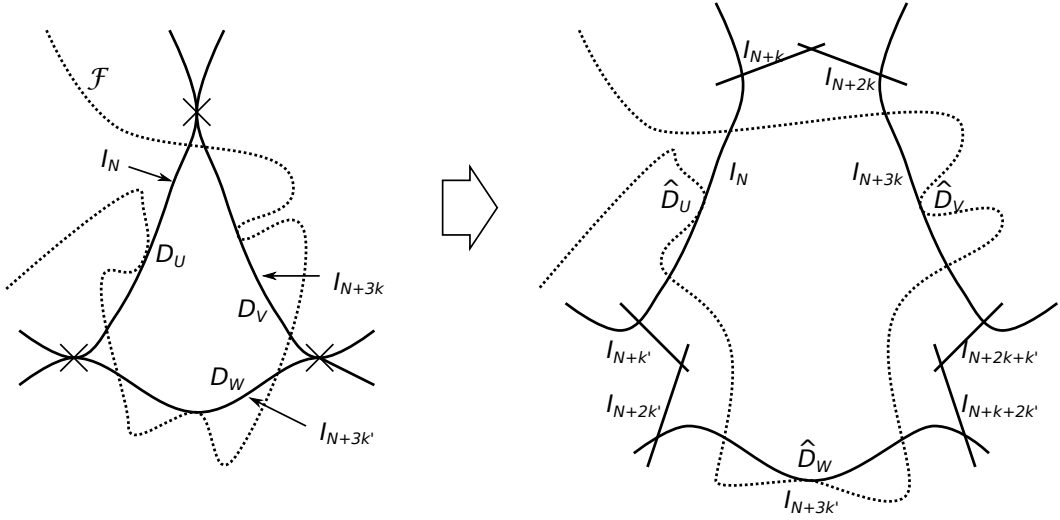}
\caption{\small A schematic diagram of the configuration
of divisors on $B$ (left) and $\hat B$ (right).
The singularity type of the elliptic fiber over each
divisor is indicated.
On the left, $D_x$ are the fractional divisors,
while the dotted line represents $\cF$,
the residual divisor of the discriminant
of the elliptic fibration.
The points where the fundamentals matter of
$G_x$ lie are represented by points where
$\cF$ meets $D_x$
transversally, and the points
where antisymmetrics lie are represented by
the points where $\cF$ meets $D_x$
tangentially. Each pair of divisors $D_x$ and $D_y$
meet at a single orbifold point, where superconformal matter jointly
charged under $G_x \times G_y$ lie.
On the right, $\hat D_x$ are integral divisors on
$\hat B$ obtained by resolving $D_x$.
The theory now has only ordinary matter.
In particular, there exist bifundamental matter
at the intersection loci of adjacent divisors.
}
\label{f:UVW}
\end{figure}

Let us now check the gauge and mixed
anomaly equations for the gauge groups.
We neglect the gravitational anomaly to avoid
further cluttering of equations, but comment on
it later on.
The contribution to the gauge and mixed anomalies
from the vector and hypermultiplets can be computed to be
\bea
I^{g,m}_{1\ell}=&-{1 \ov 16} \tr R^2 (\tr F_U^2 + \tr F_V^2 + \tr F_W^2)
-{1 \ov 8} ((\tr F_U^2)^2 + (\tr F_V^2)^2 + (\tr F_W^2)^2) \\
&+{1 \ov 96} \tr R^2 \Big\{ (2N+k+k') \tr F_U^2
+(2N+4k+k') \tr F_V^2 + (2N+k+4k') \tr F_W^2 \Big\} \\
&-{1 \ov 24} \Big\{ (2N+k+k') \tr F_U^4
+(2N+4k+k') \tr F_V^4 + (2N+k+4k') \tr F_W^4 \Big\} \,.
\eea
The field strength $F_x$ is that of the gauge symmetry
that lies above the divisor $D_x$.
Using the anomaly polynomial for the strongly
coupled SCFTs computed in \eq{anom Nkp},
we find that the total gauge and mixed anomaly
contribution is given by
\bea
I^{g,m}_S &= {1 \ov  6}
\Big\{
(\tr F_U^2)^2+(\tr F_V^2)^2+(\tr F_W^2)^2
\Big\}
+ {1 \ov 12} (\tr F_U^2 \tr F_V^2
+\tr F_V^2 \tr F_W^2 + \tr F_W^2 \tr F_U^2)
 \\
&-{1 \ov 96}\tr R^2
\Big\{ (2N+k+k') \tr F_U^2
+(2N+4k+k') \tr F_V^2 + (2N+k+4k') \tr F_W^2 \Big\} \\
&+{1 \ov 24} \Big\{ (2N+k+k') \tr F_U^4
+(2N+4k+k') \tr F_V^4 + (2N+k+4k') \tr F_W^4 \Big\}
\,.
\eea
The second and third lines for the expressions
of $I_{1 \ell}$ and $I_S$ cancel upon summing the
two contributions of the total anomaly:
\be
I^{g,m}_{1 \ell} + I^{g,m}_S
= -{1 \ov 16} \tr R^2 (\tr F_U^2 + \tr F_V^2 + \tr F_W^2)
+{1 \ov 8} \times {1 \ov 3} (\tr F_U^2 + \tr F_V^2 + \tr F_W^2)^2 \,.
\label{gmtot}
\ee
We have isolated the factor of $1/3$ to emphasize the
fractional quantization of the gauge anomaly term.
Recall that in theories with only conventional multiplets
contributing to the anomaly, the gauge anomaly
term is quantized in units of $1/8$.
The gauge and mixed anomaly terms are cancelled
by the GSSW mechanism with the gauge anomaly coefficients
of the $SU(N)$, $SU(N+3k)$ and $SU(N+3k')$
groups taken to be $D_U$, $D_V$ and $D_W$,
respectively:
\bea
I_{GS}={1 \ov 32} \Big\{
{K \ov 2} \tr R^2 +
2D_U \tr F_U^2 + 2D_V \tr F_V^2 + 2D_W \tr F_W^2
\Big\}^2 \,.
\eea
Using the intersection relations \eq{Dxint},
we find that the $\tr R^2 \tr F^2$ and $(\tr F^2)^2$
terms of $I_{GS}$ agrees precisely with equation
\eq{gmtot}. The gravitational anomaly cancellation
condition is satisfied with
\be
\nu_H = 6+ {3 \ov 2} (8-N-k-k')(7-N-k-k')
\ee
neutral hypermultiplets.

Upon moving to a generic point in
the tensor branch of all the SCFTs in the
chamber studied in section \ref{ss:B},
we can resolve the manifold $B$ to $\hat B$.
Now the effective theory is a supergravity theory
with six tensor multiplets, and a gauge group
that consists of nine special-unitary components.
The divisor configuration on $\hat B$ is depicted
on the right panel of figure \ref{f:UVW}.
The string charge of the instanton of
gauge groups $G_x$
lies in the homology lattice of this manifold---it
is given by $\hat D_x$, which have been written
out explicitly in terms of the basis
$\{ h \} \coprod \{ D_{xy,a} \}$ in \eq{DUVW}.
Notice that while we have shown that
the anomaly coefficients of $G_x$ all can be
identified with $[D_x]=h/3$, the string
charge of the $G_x$-instantons all differ.

Let us denote the gauge group with support above the
resolution divisor $D_{xy,a}$ as $G_{xy,a}$.
Then the matter content of the theory on $\hat B$
is simple to specify. The hypermultiplet matter that are
only charged under
\be
G_U \times G_V \times G_W = SU(N) \times SU(N+3k) \times SU(N+3k')
\ee
is equivalent to that of the theory on $B$.
The difference is that the superconformal
theories are gone, and that there is a bifundamental
hypermultiplet for each pair of adjacent divisors.
Using this matter content, we find that the anomaly
cancellation conditions are satisfied with the gauge
anomaly coefficients given by the homology
class of the loci of the gauge branes.
In particular, it is simple to verify that the divisors
$\hat D_x$ have self-intersection $(-1)$ while $D_{xy,a}$
have self-intersection $(-2)$,
using the anomaly cancellation conditions.
This is a simple consistency check that the
field theory computation agrees with the geometry
described in section \ref{ss:B}.

\section{Conclusions and future directions}
\label{s:future}

In this paper we have studied the question of how to couple a 6D superconformal
field theory to gravity. To accomplish this, we have studied 6D F-theory vacua
compactified on an elliptic Calabi-Yau threefold in which two-cycles of the base collapse to zero size.
In particular, we have shown that when the base has orbifold singularities, the data of the anomaly polynomial is correctly reproduced by
the intersection theory of the orbifold base. We have also seen how charge quantization predicts the existence
of additional light states -- namely those of the SCFT -- and how this can be interpreted as a refinement of the lattice
spanned by the fractional divisors of the orbifold theory. We have also presented a compact model which illustrates all of these elements.
In the remainder of this section we discuss some potential directions for future investigation.

Our primary focus has been on recoupling a particular class of 6D SCFTs to gravity. Now, one result from
\cite{HMV} is a classification of all possible orbifold singularities which a non-compact F-theory model could possess. Locally, these are
all of the form $\mathbb{C}^2 / \Gamma$ for $\Gamma$ a discrete subgroup of $U(2)$. In particular, the exact list of possible $\Gamma$'s has been determined. It would be very interesting to determine the subset of such SCFTs which can be recoupled to gravity.

One of the motivations for the present work has been to see how much of F-theory can be phrased purely in terms of the intersection theory
of fractional divisors. We anticipate that
this feature will be particularly important in the context of 4D vacua, where there may be
obstructions to motion on the geometric moduli space. Along these lines, it would be quite interesting to develop a similar analysis
for orbifolds of the form $\mathbb{C}^3 / \Gamma$ for $\Gamma$ a discrete subgroup of $U(3)$.

Finally, there is a conceptual point connected with the coupling of a 6D SCFT to gravity. On the one hand, this is straightforward to realize
using the geometry of an F-theory compactification. On the other hand, the absence of a Lagrangian description for these theories renders a purely field theoretic analysis quite subtle. It would be interesting to study in more detail how a 6D effective field theorist would infer (perhaps along the lines of \cite{Heckman:2013kza,Balasubramanian:2014bfa}) that conformal symmetry has been broken by having a finite Planck scale.

\acknowledgments{
We thank B. Haghighat, T. Rudelius, Y. Tachikawa, W. Taylor
and C. Vafa for helpful discussions. MDZ, DRM and DP also
thank the 2014 Summer Workshop at the Simons Center for Geometry and
Physics for hospitality, where some of this work was completed. %The work of JJH is supported by biscuits and gravy.
The work of MDZ is supported by NSF grant PHY-1067976. The work of DRM\ is
supported by NSF grant PHY-1307513. The work of DP is supported by
DOE grant DE-FG02-92ER-40697.
}

\appendix

\section{Anomaly polynomials of supergravity multiplets}
\label{ap:anomaly polynomials}

In this section, we review anomaly polynomials of
6D (1,0) supergravity multiplets.
We use normalization conventions of
\cite{Green:1987mn}, but with an overall minus sign.
This is to make our conventions consistent with
the anomaly calculations in the literature
\cite{Erler:1993zy,FHMM,HMM,Intriligator20,OST,OSTY,Intriligator}.
The list of ``conventional multiplets" of six-dimensional
supergravity whose anomaly polynomials
we consider is given in table \ref{t:mult}.

The anomaly polynomial of the gravity multiplet can be obtained
by summing contributions from the self-dual tensor and the
gravitino. It is given by
\bea
I_{G}
&= -{273 \ov 5760} \tr R^4 + {17 \ov 1536} (\tr R^2)^2
= -{273 \ov 5760} (\tr R^4 + {5 \ov 4} (\tr R^2)^2)+{9 \ov 128} (\tr R^2)^2 \,.
\eea
The anomaly polynomial of the tensor multiplet
can be obtained from contributions from its anti-self-dual
tensor and fermion:
\bea
I_{T}
&= {29 \ov 5760} \tr R^4 - {7 \ov 4608} (\tr R^2)^2
= {29 \ov 5760} (\tr R^4 + {5 \ov 4} (\tr R^2)^2)-{1 \ov 128} (\tr R^2)^2 \,.
\eea
Meanwhile, the contribution to the anomaly polynomials of
the vector and hypermultiplets comes solely from its fermions.
For a vector multiplet, the anomaly polynomial is given by
\bea
I_{V,G}
&=- {1 \ov 5760} (\tr R^4 + {5 \ov 4} (\tr R^2)^2) ( \tr_\text{Adj} \mathbf{1})
+{1 \ov 96} \tr R^2 \tr_\text{Adj} F^2
-{1 \ov 24} \tr_\text{Adj} F^4 \,,
\eea
while the anomaly polynomial for a hypermultiplet charged
in the representation $\bf R$ of the gauge group is given by
\bea
I_{H,{\bf R}}
&= {1 \ov 5760} (\tr R^4 + {5 \ov 4} (\tr R^2)^2)( \tr_{\bf R} \mathbf{1})
-{1 \ov 96} \tr R^2 \tr_{\bf R} F^2
+{1 \ov 24} \tr_{\bf R} F^4 \,.
\eea
We omit the subscript $R$ when taking the trace in
the fundamental representation.

\begin{table}[t!]
\center
 \begin{tabular}{ | c | c |}
 \hline
 Multiplet & Field Content\\ \hline\hline
 Gravity & $(g_{\mu \nu}, \psi^+_\mu, B^+_{\mu \nu}) $  \\ \hline
 Tensor & $(\phi, \chi^-, B^-_{\mu \nu})$   \\ \hline
 Vector & $(A_{\mu}, \lam^+)$   \\ \hline
 Hyper & $(4\varphi, \psi^-)$  \\ \hline
  \end{tabular}
 \caption{Conventional multiplets
  of 6D (1,0) supergravity theories.
  The superscripts on the fermions denote the chirality, while
  those on the antisymmetric tensors
  indicate self-duality/anti-self-duality.}
\label{t:mult}
\end{table}

\section{$SU(N)$ models on $B = \bP^2/\bZ_3$
with no charged strongly coupled sector}
\label{ap:coefficients}

In this appendix, we count the degrees of freedom
in the Weierstrass models that engineer the $SU(N)$
theories of section \ref{ss:noSCFT}. To be more precise,
we write down a generic Weierstrass model of an elliptically
fibered manifold over $B=\bP^2 / \bZ_3$ whose low
energy theory is a supergravity theory coupled to three
$(2,0)$ $A_2$ theories that has an $SU(N)$ gauge group
with the following matter content:
\be
SU(N): \;\;\;\;\; \;\;\;\;\;
(24-3N) \times {\tiny\yng(1)} +
3 \times {\tiny\yng(1,1)}+
1 \times {\rm Adj} \,,
\;\;\;\;\; [\s] = h \,.
\label{suNmodels}
\ee
The gauge and mixed anomaly cancellation involving the
$SU(N)$ gauge group for these
models have been verified in equation
\eq{suN GSSW}.
The Weierstrass model is given by
\be
y^2 =x^3 + f x +g
\ee
where $f$ and $g$ are sections of $12H$ and $18H$
of $\bP^2$ that are invariant under the $\bZ_3$
action, respectively. A section of $nH$ is represented
by a degree-$n$ homogeneous polynomial of the projective
coordinates $U$, $V$ and $W$.
The $SU(N)$ locus is represented
by the divisor $\s$ that lifts to the divisor
\be
\ts = aU^3 + bV^3 + c + d UVW \,,
\ee
of the cover $\tB=\bP^2$ of $B$.
We can follow the analysis of \cite{MTMatter}
that we have already used in section \ref{ss:noSCFT}
to count the number of Weierstrass coefficients for
$SU(2)$ models. We continue applying this
analysis to $SU(N)$ models for $3 \leq N \leq 8$
in section \ref{ss:Weierstrass}.

An interesting phenomenon can be observed in
this particular class of models in F-theory.
Given that there is
a $SU(6)$ or $SU(7)$ non-abelian gauge symmetry,
the F-theory model automatically turns out to
have abelian gauge symmetry as well.
In fact, there is an
additional $U(1)$ factor for both models.
This can be confirmed by identifying a rational
section of the elliptic fibration, and also by
successively Higgsing the $SU(8)$ model
to arrive at the $SU(6)$ and $SU(7)$ model.
We discuss issues related to the abelian
gauge symmetry of the theory in section
\ref{ss:MW}.

\subsection{Complex degrees of freedom in
Weierstrass models}
\label{ss:Weierstrass}

The number of complex degrees of freedom of the
Weierstrass
models can be systematically computed by expanding
the Weierstrass coefficients $f$ and $g$ with respect to
the $SU(N)$ locus $\ts$ \cite{MTMatter}:
\be
f= \sum_i f_i \ts^i, \quad g = \sum_i g_i \ts^i \,.
\ee
Then, the coefficients $\Delta_i$ of the
discriminant locus $\Delta$ of the model
in $\ts$,
\be
\Delta = 4f^3 + 27 g^2 = \sum_i \Delta_i \ts^i \,,
\ee
can be written in terms of $f_i$ and $g_i$.
Imposing that $\Delta$ has vanishing coefficients
up to order $\ts^{N-1}$, we obtain constraints on
the coefficients $f_i$ and $g_i$. The degrees of
freedom of the Weierstrass model is obtained by
counting by the number of degrees of
freedom in the solutions of these constraints.
\vspace*{0.05in}

\noindent
\textbf{SU(3) : 38 complex degrees of freedom}

\noindent
For an $SU(3)$ theory,
the Weierstrass coefficients $f$ and $g$ must
be of the form
\bea
f &= -{1 \ov 48} \phi_0^4 + {1 \ov 2} \phi_0 \psi_1 \ts
+f_2 \ts^2 + f_3 \ts^3 \\
g &= {1 \ov 864} \phi_0^6 - {1 \ov 24} \phi_0^3 \psi_1 \ts
+\left( {1 \ov 4} \psi_1^2 -{1 \ov 12} \phi_0^2 f_2 \right) \ts^2
+ g_3 \ts^3
\label{su3}
\eea
$\phi_0$, $\psi_1$, $f_2$, $f_3$ and $g_3$
are sections of $3H$, $6H$, $6H$, $3H$ and $9H$,
respectively. While $f_3$ and $g_3$ are generic sections,
the other degree-$n$ homogeneous polynomials
of the projective coordinates
can always be reduced to the form,
\be
W^n ( p_0 (v^3) + p_1 (v^3) uv+  p_2(v^3) u^2v^2)
\label{rf}
\ee
where $u = U/W$, $v=V/W$,
and $p_{0,1,2}$ are polynomials.
We denote any polynomial of the form \eq{rf}
to be in a ``reduced form."
This is due to the fact that these sections are
$\bZ_3$ invariant, and that any factor of $u^3$
can be replaced using the divisor $\ts$.
For example, $\phi_0$ can be written as
\be
\phi_0 = W^3 (\varphi_{0,0} (v^3) + \varphi_1 uv )
\ee
which has three complex degrees of freedom.
Meanwhile, a generic $\bZ_3$ invariant section
of $3nH$ has
\be
\binom{n+2}{2}+\binom{n+1}{2}+\binom{n}{2}
\ee
complex degrees of freedom. Summing up all the complex
degrees of freedom available, including those in $\ts$,
we see that there are
$42$ free coefficients in \eq{su3}.

We must now subtract the number of symmetries and
automorphisms available to arrive at the number of
complex degrees of freedom of the Weierstrass model.
On top of the $(\bC^*)^3$ automorphism, there exists
the $\bC^*$ symmetry
\be
\ts \ra t \ts, \quad
\psi_1 \ra t^{-1} \psi_1, \quad
f_n \ra t^{-n} f_n, \quad
g_n \ra t^{-n} g_n \,.
\ee
of the Weierstrass model.
The final number of complex degrees of freedom
is given by $42-4 =38$.

\vspace*{0.05in}

\noindent
\textbf{SU(4) : 26 complex degrees of freedom}

\noindent
The Weierstrass coefficients $f$ and $g$ must
be of the form
\bea
f &= -{1 \ov 48} \phi_0^4 - {1 \ov 6} \phi_0^2 \phi_1 \ts
+f_2 \ts^2 + f_3 \ts^3 + f_4 \ts^4 \\
g &= {1 \ov 864} \phi_0^6 +{1 \ov 72} \phi_0^4 \phi_1 \ts
+\left( {1 \ov 36} \phi_0^2 \phi_1^2 -{1 \ov 12} \phi_0^2 f_2 \right) \ts^2
+ \left( -{1 \ov 12} \phi_0^2 f_3 -{1 \ov 3} \phi_1 f_2 -{1 \ov 27} \phi_1^3 \right) \ts^3
+ g_4 \ts^4 \,.
\label{su4}
\eea
$\phi_0$, $\phi_1$, $f_2$ and $f_3$
are sections of $3H$, $3H$, $6H$ and $3H$
in reduced form, while $f_4$ is a constant,
and $g_4$ is a generic section of $6H$.
The total number of complex coefficients
in \eq{su4} is 30.
Meanwhile, in addition to
the $(\bC^*)^3$ automorphism, there exists
the $\bC^*$ symmetry
\be
\ts \ra t \ts, \quad
\phi_1 \ra t^{-1} \phi_1, \quad
f_n \ra t^{-n} f_n, \quad
g_n \ra t^{-n} g_n \,.
\ee
of the Weierstrass model.
The number of complex degrees of freedom
is given by $30-4 =26$.
\vspace*{0.05in}

\noindent
\textbf{SU(5) : 17 complex degrees of freedom}

\noindent
The Weierstrass coefficients $f$ and $g$ must
be of the form
\bea
f &= -{1 \ov 48} \phi_0^4 - {1 \ov 6} \phi_0^2 \phi_1 \ts
+\left( {1 \ov 2} \phi_0 \psi_2 -{1 \ov 3} \phi_1^2\right) \ts^2
+ f_3 \ts^3 + f_4 \ts^4 \\
g &= {1 \ov 864} \phi_0^6 +{1 \ov 72} \phi_0^4 \phi_1 \ts
+\left( {1 \ov 18} \phi_0^2 \phi_1^2 -{1 \ov 24} \phi_0^3 \psi_2 \right) \ts^2 \\
&+ \left( -{1 \ov 12} \phi_0^2 f_3 -{1 \ov 6} \phi_0 \phi_1 \psi_2
+{2 \ov 27} \phi_1^3 \right) \ts^3
+\left( {1 \ov 4} \psi_2^2 -{1 \ov 12} \phi_0^2 f_4 -{1 \ov 3} \phi_1 f_3 \right) \ts^4
+ g_5 \ts^5 \,.
\label{su5}
\eea
$\phi_0$, $\phi_1$, $\psi_2$ and $f_3$
are sections of $3H$ of reduced form,
while $f_4$ is a constant
and $g_5$ is a generic section of $3H$.
The total number of complex coefficients
in \eq{su5} is 21.
As before, in addition to
the $(\bC^*)^3$ automorphism, there is
a $\bC^*$ symmetry
\be
\ts \ra t \ts, \quad
\phi_1 \ra t^{-1} \phi_1, \quad
\psi_2 \ra t^{-2} \psi_2,\quad
f_n \ra t^{-n} f_n, \quad
g_n \ra t^{-n} g_n \,.
\ee
of the Weierstrass model.
The number of complex degrees of freedom
is given by $21-4 =17$.
\vspace*{0.05in}

\noindent
\textbf{SU(6) : 12 complex degrees of freedom}

\noindent
The Weierstrass coefficients $f$ and $g$ must
be of the form
\bea
f &= -{\beta^4 \ov 48}   \alpha^4  - {\beta^3 \ov 6}    \alpha^2\nu \ts
+\left( -{\beta \phi_2 \ov 6}   \alpha^2 -{\beta^2 \ov 3}  \nu^2 \right) \ts^2
+ \left(  - (3 \beta) \lambda-{\phi_2 \ov 3}  \nu  \right) \ts^3 + f_4 \ts^4 \\
g &= {\beta^6 \ov 864}  \alpha^6  +{\beta^5 \ov 72}  \alpha^4  \nu \ts
+\left( { \beta^3 \phi_2 \ov 72} \alpha^4  +{ \beta^4 \ov 18} \alpha^2  \nu^2  \right) \ts^2 \\
&+ \left( {\beta^3 \ov 4}  \alpha^2  \lambda +{\beta^2 \phi_2 \ov 12}  \alpha^2 \nu
+{2  \beta^3 \ov 27} \nu^3 \right) \ts^3 \\
&+\left( \left( {1 \ov 36} \phi_2^2  -{1 \ov 12} \beta^2 f_4 \right) \alpha^2
+  (\beta^2) \lambda \nu   +{\beta \phi_2 \ov 9}  \nu^2  \right) \ts^4
+\left( (\phi_2) \lambda -{\beta f_4 \ov 3}  \nu  \right) \ts^5
+ g_6 \ts^6 \,.
\label{su6}
\eea
$\alpha$, $\nu$, $\lambda$
are sections of $3H$ of reduced form,
while $\beta$, $\phi_2$, $f_4$ and $g_6$
are constants.
The total number of complex coefficients
in \eq{su6} is 17.
On top of
the $(\bC^*)^3$ automorphism, there is
a $(\bC^*)^2$ symmetry
\bea
\ts \ra t \ts, \quad
\nu \ra t^{-1} \nu, \quad
\phi_2 \ra t^{-2} \phi_2,\quad
\lambda \ra t^{-3} \lambda,\quad
f_4 \ra t^{-4} f_4, \quad
g_6 \ra t^{-6} g_6 \\
\alpha \ra s \alpha, \quad
\beta \ra s^{-1} \beta, \quad
\nu \ra s \nu, \quad
\lambda \ra s \lambda, \quad
\phi_2 \ra s^{-1} \phi_2
\label{cstar2}
\eea
of the Weierstrass model.
The number of complex degrees of freedom
is given by $17-5 =12$.
\vspace*{0.05in}

\noindent
\textbf{SU(7) : 9 complex degrees of freedom}

\noindent
The Weierstrass coefficients $f$ and $g$ must
be of the form \eq{su6} with
\bea
\lambda = {1 \ov 9\beta} \nu \phi_2 -{1 \ov 6} \psi_3 \alpha , \quad
g_6 = -{1 \ov 27\beta^3} \phi_2^3 + {1 \ov 4} \psi_3^2 -{1 \ov 3 \beta} \phi_2 f_4 \,.
\label{su7}
\eea
Now the moving pieces are $\alpha$ and $\nu$,
which are sections of $3H$ in reduced form,
and the constants $\beta$, $\phi_2$, $\psi_3$ and $f_4$,
and of course, the brane locus $\ts$.
The total number of complex coefficients
is given by \eq{su6} is 14.
As usual, there is a
the $(\bC^*)^3$ automorphism,
and the $(\bC^*)^2$ symmetry
\bea
\ts \ra t \ts, \quad
\nu \ra t^{-1} \nu, \quad
\phi_2 \ra t^{-2} \phi_2,\quad
\psi_3 \ra t^{-3} \psi_3,\quad
f_4 \ra t^{-4} f_4 \,, \\
\alpha \ra s \alpha, \quad
\beta \ra s^{-1} \beta, \quad
\nu \ra s \nu, \quad
\lambda \ra s \lambda, \quad
\phi_2 \ra s^{-1} \phi_2
\eea
of the Weierstrass model.
The number of complex degrees of freedom
is given by $14-5 =9$.
\vspace*{0.05in}

\noindent
\textbf{SU(8) : 8 complex degrees of freedom}

\noindent
The $SU(8)$ theory can be obtained from
the $SU(7)$ theory by taking
\be
\psi_3 = 0 \,.
\ee
Therefore the complex
degrees of freedom of this theory
is one less than that of the $SU(7)$
theory, which is $9-1 =8$.

The $SU(8)$ model is maximal---it is
the model with maximum
rank in the family of models \eq{suNmodels}.
The Weierstrass coefficient for such
models is given by \cite{MTMatter}
\bea
f&= -{1 \ov 3} \Phi^2 + F_4 \ts^4 \\
g&= {2 \ov 27} \Phi^3 -{1 \ov 3} F_4 \Phi \ts^4 \\
\Phi &= {1 \ov 4} \phi_0^2 + \phi_1 \ts + \phi_2 \ts^2 \,.
\label{su8}
\eea
It is simple to see verify that this is indeed the case
with the identification
\be
\Phi = {\alpha^2 \beta^2 \ov 4} + \beta \nu \ts + {\phi_2 \ov \beta} \ts^2 \,,
\qquad
F_4 = f_4+ {\phi_2^2 \ov 3 \beta^2}  \,.
\ee

\subsection{Abelian factors and the Mordell-Weil group}
\label{ss:MW}

The Weierstrass models of the theories with
$SU(6)$ \eq{su6} and $SU(7)$ \eq{su7}
gauge symmetry both have
a non-trivial Mordell-Weil group of rank one.
This implies that the theory has a $U(1)$ gauge group
in addition to the non-abelian $SU(N)$ group.
This can come as a surprise from the way we
have arrived at the models \eq{su6} and \eq{su7}.
We have successively tuned the Weierstrass coefficients
to get higher-rank non-abelian gauge symmetry,
and have not aimed at producing an elliptic fibration
with a rational section.

From the top-down point of view, however, the existence
of the $U(1)$ factor is inevitable.
The family of models \eq{suNmodels} in F-theory are
obtained from Higgsing the adjoint hypermultiplet
of the $SU(8)$ theory. When we Higgs the
$SU(8)$ theory using the adjoint field and preserve an
$SU(6)$ or $SU(7)$ gauge symmetry, there always exists
a $U(1)$ subgroup whose charged matter all are in
non-trivial representations of the non-abelian gauge
group. This $U(1)$ subgroup
thus cannot be Higgsed away without
disrupting the non-abelian gauge symmetry, and
remains unbroken.

Let us first verify the existence of a rational section
for the $SU(6)$ and $SU(7)$ models.
Recall that the Weierstrass model of an elliptic fibration
over $\bP^2$
with a rational section can be written in the form
\cite{MP}
\bea
y^2&=x^3+(2f_{3+n}f_{9-n}-3f_6^2-b_n^2 f_{12-2n})x \\
&+(2f_6^3-2f_{3+n}f_6f_{9-n}+f_{3+n}^2 f_{12-2n}-2b_n^2f_6 f_{12-2n} +{b_n^2 f_{9-n}^2}) \,,
\label{MW1}
\eea
where $b_n$ is a section of $nH$ while
$f_k$ are sections of $kH$.\footnote{Elliptic fibrations
with non-zero Mordell-Weil rank have recently been used
to construct phenomenologically interesting F-theory models.
A small sample of such work is collected in the bibliography
\cite{Mayrhofer:2012zy,Braun:2013yti,Cvetic:2013nia,
Antoniadis:2014jma,Esole:2014dea}.}
A rational section of this elliptic fibration
is then given by
\be
x= {f_{3+n}^2 \ov b_n^2} -2f_6, \quad
y=-{f_{3+n}^3 \ov b_n^3} +3 {f_6 f_{3+n} \ov b_n}-b_n f_{9-n} \,.
\ee
For the $SU(6)$ and $SU(7)$ models, $n=0$
in equation \eq{MW1} and $f_k$ and $b_0$ are given by
\bea
b_0 &= \sqrt{G_6}, \quad
f_3 = -{3 \ov 2} \Lambda, \quad
f_6 = {1 \ov 3} \Phi \,, \\
f_9& = \ts^3 - {3 \ov G_6} \Phi \Lambda
+{27 \ov 4 G_6^2} \Lambda^3 \,, \qquad
f_{12} = -{3 \ov G_6} \Lambda
\left( 2 \ts^3
-{3 \ov G_6} \Phi \Lambda
+{27 \ov 4G_6 ^2} \Lambda^3
\right) \,.
\eea
Here, $\Lambda$, $\Phi$ and $G_6$ are related
to the sections in equation \eq{su6} by
\bea
\Phi &= {\alpha^2 \beta^2 \ov 4} + \beta \nu \ts + {\phi_2 \ov \beta} \ts^2 \,, \\
\Lambda &= {1 \ov 9} \nu \phi_2 - \beta \lambda +
\left( {\phi_2^2 \ov 3 \beta^2} + f_4 \right) \ts \,,\quad
G_6 = {1 \ov 4}
\left( g_6 + {f_4 \phi_2 \ov 3 \beta} + {\phi_2^3 \ov 27 \beta^3}
\right) \,.
\eea
The $SU(7)$ model is obtained from the $SU(6)$
model by tuning
\be
G_6 = {1 \ov 4} \psi_3^2 \,, \quad
\Lambda = {1 \ov 2} \psi_3 \beta \alpha +
\left( f_4 + {\phi_2^2 \ov 3 \beta^2}
\right) \ts \,.
\ee
Note that in the $SU(7)$ model
\be
b_0 = {1 \ov 2} \psi_3 \,.
\ee
Hence, by taking $\psi_3 \ra 0$, we can enhance
the abelian factor present in the theory \cite{MP}.
But by taking this limit, we precisely arrive at
the $SU(8)$ model \eq{su8}. This is consistent
with the fact that
the $SU(7)$ model is obtained from the $SU(8)$
model by Higgsing the adjoint hypermultiplet.
The vacuum expectation value of the Higgs field
can be identified with the parameter $\psi_3$.

Counting the number of complex degrees of
freedom of the Weierstrass model and comparing with
the expected number of neutral hypermultiplets (table
\ref{t:coefficients}), it is clear that the $U(1)$ factors
only exist for the $SU(6)$ and $SU(7)$ models.
This has a simple explanation in terms of Higgsing from
the $SU(8)$ model \eq{su8}.
As noted before,
the $SU(7)$ \eq{su7} and $SU(6)$ \eq{su6} models are obtained
from the $SU(8)$ model by giving a vacuum expectation
value to the adjoint hypermultiplet
of the theory. The Higgsing that preserves the $SU(7)$
symmetry breaks the gauge group
down to $SU(7) \times U(1)$ rather than $SU(7)$:
\be
SU(7) \times U(1) : \;\;\;\;\; \;\;\;\;\;
3 \times ({\tiny\yng(1)},-6) +
3 \times ({\tiny\yng(1,1)},2)+
1 \times ({\rm Adj},0) \,.
\ee
The $U(1)$ factor is represented by the Cartan matrix
\be
\rm diag (1,1,1,1,1,1,1,-7)
\ee
in the fundamental representation of $SU(8)$.
Now there is no way we can break the $U(1)$
gauge symmetry further without breaking the $SU(7)$
symmetry, as all the matter charged under $U(1)$
is also charged under $SU(7)$. The F-theory model
is hence forced to have an additional abelian
gauge group factor.

Higgsing the theory further down using the
adjoint hypermultiplet to preserve
an $SU(6)$ symmetry, the gauge group is
broken down to $SU(6) \times U(1)^2$ where the
charged matter is given by
\bea
SU(6) \times U(1)_1 \times U(1)_2 : \;\;\;\;\;
3 \times (\cdot,0,-6) +
3 \times ({\tiny\yng(1)},&-1,-2) +
3 \times ({\tiny\yng(1)},1,-2)\\
&+3 \times ({\tiny\yng(1,1)},0,2)+
1 \times ({\rm Adj},0,0) \,.
\label{su6u12}
\eea
Here the $U(1)$ factors are represented by the
following two Cartan matrices of $SU(8)$:
\be
\rm diag (0,0,0,0,0,0,1,-1), \qquad
\rm diag (1,1,1,1,1,1,-3,-3) \,.
\ee
It can be seen that in model \eq{su6u12}
there exist hypermultiplets neutral under
$SU(6)$ but charged under $U(1)_2$
that can be used to Higgs away this
abelian gauge symmetry.
The gauge group $U(1)_1$, however, cannot
be broken without breaking $SU(6)$.
The phenomenon that the $SU(6)$ model \eq{su6}
automatically has a rational section is a reflection
of this fact.

Upon further Higgsing, we find that there are
always enough hypermultiplets neutral
under the non-abelian gauge group that can be
used to Higgs away all the abelian gauge group
factors. Hence a generic F-theory model \eq{suNmodels}
with $N \leq 5$ does not have a non-trivial abelian
gauge group.

\bibliographystyle{JHEP}
\bibliography{SingularBase}

\providecommand{\href}[2]{#2}\begingroup\raggedright\begin{thebibliography}{10}

\bibitem{Vafa}
C.~Vafa, {\it {Evidence for F theory}},  {\em Nucl.Phys.} {\bf B469} (1996)
  403--418, [\href{http://xxx.lanl.gov/abs/hep-th/9602022}{{\tt
  hep-th/9602022}}].

\bibitem{MV1}
D.~R. Morrison and C.~Vafa, {\it {Compactifications of F theory on Calabi-Yau
  threefolds. 1}},  {\em Nucl.Phys.} {\bf B473} (1996) 74--92,
  [\href{http://xxx.lanl.gov/abs/hep-th/9602114}{{\tt hep-th/9602114}}].

\bibitem{MV2}
D.~R. Morrison and C.~Vafa, {\it {Compactifications of F theory on Calabi-Yau
  threefolds. 2.}},  {\em Nucl.Phys.} {\bf B476} (1996) 437--469,
  [\href{http://xxx.lanl.gov/abs/hep-th/9603161}{{\tt hep-th/9603161}}].

\bibitem{TaylorTASI}
W.~Taylor, {\it {TASI Lectures on Supergravity and String Vacua in Various
  Dimensions}},  \href{http://xxx.lanl.gov/abs/1104.2051}{{\tt
  arXiv:1104.2051}}.

\bibitem{Witten:1995ex}
E.~Witten, {\it {String theory dynamics in various dimensions}},  {\em Nucl.
  Phys.} {\bf B443} (1995) 85--126,
  [\href{http://xxx.lanl.gov/abs/hep-th/9503124}{{\tt hep-th/9503124}}].

\bibitem{Witten:1995zh}
E.~Witten, {\it {Some comments on string dynamics}},
  \href{http://xxx.lanl.gov/abs/hep-th/9507121}{{\tt hep-th/9507121}}.

\bibitem{WittenSmall}
E.~Witten, {\it {Small Instantons in String Theory}},  {\em Nucl. Phys.} {\bf
  B460} (1996) 541--559, [\href{http://xxx.lanl.gov/abs/hep-th/9511030}{{\tt
  hep-th/9511030}}].

\bibitem{Strominger:1995ac}
A.~Strominger, {\it {Open p-branes}},  {\em Phys.Lett.} {\bf B383} (1996)
  44--47, [\href{http://xxx.lanl.gov/abs/hep-th/9512059}{{\tt
  hep-th/9512059}}].

\bibitem{Ganor:1996mu}
O.~J. Ganor and A.~Hanany, {\it {Small $E_8$ instantons and Tensionless Non
  Critical Strings}},  {\em Nucl. Phys.} {\bf B474} (1996) 122--140,
  [\href{http://xxx.lanl.gov/abs/hep-th/9602120}{{\tt hep-th/9602120}}].

\bibitem{SW6}
N.~Seiberg and E.~Witten, {\it {Comments on string dynamics in
  six-dimensions}},  {\em Nucl.Phys.} {\bf B471} (1996) 121--134,
  [\href{http://xxx.lanl.gov/abs/hep-th/9603003}{{\tt hep-th/9603003}}].

\bibitem{Bershadsky:1996nu}
M.~Bershadsky and A.~Johansen, {\it {Colliding singularities in F-theory and
  phase transitions}},  {\em Nucl. Phys.} {\bf B489} (1997) 122--138,
  [\href{http://xxx.lanl.gov/abs/hep-th/9610111}{{\tt hep-th/9610111}}].

\bibitem{Blum:1997fw}
J.~D. Blum and K.~A. Intriligator, {\it Consistency conditions for branes at
  orbifold singularities},  {\em Nucl. Phys. B} {\bf 506} (1997) 223--235,
  [\href{http://xxx.lanl.gov/abs/hep-th/9705030}{{\tt hep-th/9705030}}].

\bibitem{instK3}
P.~S. Aspinwall and D.~R. Morrison, {\it Point-like instantons on {K3}
  orbifolds},  {\em Nucl. Phys. B} {\bf 503} (1997) 533--564,
  [\href{http://xxx.lanl.gov/abs/hep-th/9705104}{{\tt hep-th/9705104}}].

\bibitem{Intriligator:1997dh}
K.~A. Intriligator, {\it New string theories in six-dimensions via branes at
  orbifold singularities},  {\em Adv. Theor. Math. Phys.} {\bf 1} (1998)
  271--282, [\href{http://xxx.lanl.gov/abs/hep-th/9708117}{{\tt
  hep-th/9708117}}].

\bibitem{Brunner:1997gf}
I.~Brunner and A.~Karch, {\it {Branes at orbifolds versus Hanany Witten in
  six-dimensions}},  {\em JHEP} {\bf 9803} (1998) 003,
  [\href{http://xxx.lanl.gov/abs/hep-th/9712143}{{\tt hep-th/9712143}}].

\bibitem{Hanany:1997gh}
A.~Hanany and A.~Zaffaroni, {\it {Branes and six-dimensional supersymmetric
  theories}},  {\em Nucl.Phys.} {\bf B529} (1998) 180--206,
  [\href{http://xxx.lanl.gov/abs/hep-th/9712145}{{\tt hep-th/9712145}}].

\bibitem{HMV}
J.~J. Heckman, D.~R. Morrison, and C.~Vafa, {\it {On the Classification of 6D
  SCFTs and Generalized ADE Orbifolds}},  {\em JHEP} {\bf 1405} (2014) 028,
  [\href{http://xxx.lanl.gov/abs/1312.5746}{{\tt arXiv:1312.5746}}].

\bibitem{DHTV}
M.~Del~Zotto, J.~J. Heckman, A.~Tomasiello, and C.~Vafa, {\it {6d Conformal
  Matter}},  \href{http://xxx.lanl.gov/abs/1407.6359}{{\tt arXiv:1407.6359}}.

\bibitem{Heckman}
J.~J. Heckman, {\it {More on the Matter of 6D SCFTs}},
  \href{http://xxx.lanl.gov/abs/1408.0006}{{\tt arXiv:1408.0006}}.

\bibitem{sixTOTAL}
J.~J. Heckman, D.~R. Morrison, T.~Rudelius, and C.~Vafa, {\it {To Appear}}, .

\bibitem{KMT2}
V.~Kumar, D.~R. Morrison, and W.~Taylor, {\it {Global aspects of the space of
  6D N = 1 supergravities}},  {\em JHEP} {\bf 1011} (2010) 118,
  [\href{http://xxx.lanl.gov/abs/1008.1062}{{\tt arXiv:1008.1062}}].

\bibitem{SeibergTaylor}
N.~Seiberg and W.~Taylor, {\it {Charge Lattices and Consistency of 6D
  Supergravity}},  {\em JHEP} {\bf 1106} (2011) 001,
  [\href{http://xxx.lanl.gov/abs/1103.0019}{{\tt arXiv:1103.0019}}].

\bibitem{Cordova}
C.~Cordova, {\it {Decoupling Gravity in F-Theory}},  {\em Adv.Theor.Math.Phys.}
  {\bf 15} (2011) 689--740, [\href{http://xxx.lanl.gov/abs/0910.2955}{{\tt
  arXiv:0910.2955}}].

\bibitem{FUZZ}
J.~J. Heckman and H.~Verlinde, {\it {Evidence for F(uzz) Theory}},  {\em JHEP}
  {\bf 1101} (2011) 044, [\href{http://xxx.lanl.gov/abs/1005.3033}{{\tt
  arXiv:1005.3033}}].

\bibitem{BanksSeiberg}
T.~Banks and N.~Seiberg, {\it {Symmetries and Strings in Field Theory and
  Gravity}},  {\em Phys.Rev.} {\bf D83} (2011) 084019,
  [\href{http://xxx.lanl.gov/abs/1011.5120}{{\tt arXiv:1011.5120}}].

\bibitem{milnor}
J.~Milnor, {\it {On simply connected 4-manifolds}},  {\em Symposio
  internacional de topologia algebrica, UNAM and UNESCO, Mexico City} (1958)
  pp. 122--128.

\bibitem{Green:1984bx}
M.~B. Green, J.~H. Schwarz, and P.~C. West, {\it Anomaly free chiral theories
  in six-dimensions},  {\em Nucl. Phys. B} {\bf 254} (1985) 327--348.

\bibitem{Sagnotti1}
A.~Sagnotti, {\it {A Note on the Green-Schwarz mechanism in open string
  theories}},  {\em Phys.Lett.} {\bf B294} (1992) 196--203,
  [\href{http://xxx.lanl.gov/abs/hep-th/9210127}{{\tt hep-th/9210127}}].

\bibitem{Sadov:1996zm}
V.~Sadov, {\it Generalized {G}reen--{S}chwarz mechanism in {F}-theory},  {\em
  Phys. Lett. B} {\bf 388} (1996) 45--50,
  [\href{http://xxx.lanl.gov/abs/hep-th/9606008}{{\tt hep-th/9606008}}].

\bibitem{anomalies}
A.~Grassi and D.~R. Morrison, {\it Anomalies and the {E}uler characteristic of
  elliptic {C}alabi--{Y}au threefolds},  {\em Commun. Number Theory Phys.} {\bf
  6} (2012), no.~1 51--127, [\href{http://xxx.lanl.gov/abs/1109.0042}{{\tt
  arXiv:1109.0042}}].

\bibitem{MR0459426}
C.~W. Bernard, N.~H. Christ, A.~H. Guth, and E.~J. Weinberg, {\it
  Pseudoparticle parameters for arbitrary gauge groups},  {\em Phys. Rev. D
  (3)} {\bf 16} (1977), no.~10 2967--2977.

\bibitem{OSTY}
K.~Ohmori, H.~Shimizu, Y.~Tachikawa, and K.~Yonekura, {\it {Anomaly polynomial
  of general 6d SCFTs}},  \href{http://xxx.lanl.gov/abs/1408.5572}{{\tt
  arXiv:1408.5572}}.

\bibitem{Erler:1993zy}
J.~Erler, {\it Anomaly cancellation in six dimensions},  {\em J. Math. Phys.}
  {\bf 35} (1994) 1819--1833,
  [\href{http://xxx.lanl.gov/abs/hep-th/9304104}{{\tt hep-th/9304104}}].

\bibitem{FMS}
S.~Ferrara, R.~Minasian, and A.~Sagnotti, {\it {Low-energy analysis of M and F
  theories on Calabi-Yau threefolds}},  {\em Nucl.Phys.} {\bf B474} (1996)
  323--342, [\href{http://xxx.lanl.gov/abs/hep-th/9604097}{{\tt
  hep-th/9604097}}].

\bibitem{KMT1}
V.~Kumar, D.~R. Morrison, and W.~Taylor, {\it {Mapping 6D N = 1 supergravities
  to F-theory}},  {\em JHEP} {\bf 1002} (2010) 099,
  [\href{http://xxx.lanl.gov/abs/0911.3393}{{\tt arXiv:0911.3393}}].

\bibitem{HellermanSharpe}
S.~Hellerman and E.~Sharpe, {\it {Sums over topological sectors and
  quantization of Fayet-Iliopoulos parameters}},  {\em Adv.Theor.Math.Phys.}
  {\bf 15} (2011) 1141--1199, [\href{http://xxx.lanl.gov/abs/1012.5999}{{\tt
  arXiv:1012.5999}}].

\bibitem{Deser:1997se}
S.~Deser, A.~Gomberoff, M.~Henneaux, and C.~Teitelboim, {\it {P-brane dyons and
  electric magnetic duality}},  {\em Nucl.Phys.} {\bf B520} (1998) 179--204,
  [\href{http://xxx.lanl.gov/abs/hep-th/9712189}{{\tt hep-th/9712189}}].

\bibitem{Witten:1996hc}
E.~Witten, {\it {Five-brane effective action in M theory}},  {\em J.Geom.Phys.}
  {\bf 22} (1997) 103--133, [\href{http://xxx.lanl.gov/abs/hep-th/9610234}{{\tt
  hep-th/9610234}}].

\bibitem{Aharony:1998qu}
O.~Aharony and E.~Witten, {\it {Anti-de Sitter space and the center of the
  gauge group}},  {\em JHEP} {\bf 9811} (1998) 018,
  [\href{http://xxx.lanl.gov/abs/hep-th/9807205}{{\tt hep-th/9807205}}].

\bibitem{Witten:1998wy}
E.~Witten, {\it {AdS / CFT correspondence and topological field theory}},  {\em
  JHEP} {\bf 9812} (1998) 012,
  [\href{http://xxx.lanl.gov/abs/hep-th/9812012}{{\tt hep-th/9812012}}].

\bibitem{Moore:2004jv}
G.~W. Moore, {\it {Anomalies, Gauss laws, and Page charges in M-theory}},  {\em
  Comptes Rendus Physique} {\bf 6} (2005) 251--259,
  [\href{http://xxx.lanl.gov/abs/hep-th/0409158}{{\tt hep-th/0409158}}].

\bibitem{Belov:2006jd}
D.~Belov and G.~W. Moore, {\it {Holographic Action for the Self-Dual Field}},
  \href{http://xxx.lanl.gov/abs/hep-th/0605038}{{\tt hep-th/0605038}}.

\bibitem{Freed:2006yc}
D.~S. Freed, G.~W. Moore, and G.~Segal, {\it {Heisenberg Groups and
  Noncommutative Fluxes}},  {\em Annals Phys.} {\bf 322} (2007) 236--285,
  [\href{http://xxx.lanl.gov/abs/hep-th/0605200}{{\tt hep-th/0605200}}].

\bibitem{Witten:2007ct}
E.~Witten, {\it {Conformal Field Theory In Four And Six Dimensions}},
  \href{http://xxx.lanl.gov/abs/0712.0157}{{\tt arXiv:0712.0157}}.

\bibitem{Witten:2009at}
E.~Witten, {\it {Geometric Langlands From Six Dimensions}},
  \href{http://xxx.lanl.gov/abs/0905.2720}{{\tt arXiv:0905.2720}}.

\bibitem{Henningson:2010rc}
M.~Henningson, {\it {The partition bundle of type $A_{N-1}$ (2, 0) theory}},
  {\em JHEP} {\bf 1104} (2011) 090,
  [\href{http://xxx.lanl.gov/abs/1012.4299}{{\tt arXiv:1012.4299}}].

\bibitem{Freed:2012bs}
D.~S. Freed and C.~Teleman, {\it {Relative quantum field theory}},  {\em
  Commun.Math.Phys.} {\bf 326} (2014) 459--476,
  [\href{http://xxx.lanl.gov/abs/1212.1692}{{\tt arXiv:1212.1692}}].

\bibitem{BG}
C.~P. Boyer and K.~Galicki, {\em Sasakian geometry}.
\newblock Oxford Mathematical Monographs. Oxford University Press, Oxford,
  2008.

\bibitem{OST}
K.~Ohmori, H.~Shimizu, and Y.~Tachikawa, {\it {Anomaly polynomial of E-string
  theories}},  \href{http://xxx.lanl.gov/abs/1404.3887}{{\tt arXiv:1404.3887}}.

\bibitem{Intriligator}
K.~Intriligator, {\it {6d, N=(1,0) Coulomb Branch Anomaly Matching}},
  \href{http://xxx.lanl.gov/abs/1408.6745}{{\tt arXiv:1408.6745}}.

\bibitem{GaTom}
D.~Gaiotto and A.~Tomasiello, {\it {Holography for (1,0) theories in six
  dimensions}},  {\em JHEP} {\bf 1412} (2014) 003,
  [\href{http://xxx.lanl.gov/abs/1404.0711}{{\tt arXiv:1404.0711}}].

\bibitem{FHMM}
D.~Freed, J.~A. Harvey, R.~Minasian, and G.~W. Moore, {\it {Gravitational
  anomaly cancellation for M theory five-branes}},  {\em Adv.Theor.Math.Phys.}
  {\bf 2} (1998) 601--618, [\href{http://xxx.lanl.gov/abs/hep-th/9803205}{{\tt
  hep-th/9803205}}].

\bibitem{HMM}
J.~A. Harvey, R.~Minasian, and G.~W. Moore, {\it {NonAbelian tensor multiplet
  anomalies}},  {\em JHEP} {\bf 9809} (1998) 004,
  [\href{http://xxx.lanl.gov/abs/hep-th/9808060}{{\tt hep-th/9808060}}].

\bibitem{WittenPT}
E.~Witten, {\it {Phase transitions in M theory and F theory}},  {\em
  Nucl.Phys.} {\bf B471} (1996) 195--216,
  [\href{http://xxx.lanl.gov/abs/hep-th/9603150}{{\tt hep-th/9603150}}].

\bibitem{KatzMorrisonPlesser}
S.~H. Katz, D.~R. Morrison, and M.~R. Plesser, {\it {Enhanced gauge symmetry in
  type II string theory}},  {\em Nucl.Phys.} {\bf B477} (1996) 105--140,
  [\href{http://xxx.lanl.gov/abs/hep-th/9601108}{{\tt hep-th/9601108}}].

\bibitem{MTMatter}
D.~R. Morrison and W.~Taylor, {\it {Matter and singularities}},  {\em JHEP}
  {\bf 1201} (2012) 022, [\href{http://xxx.lanl.gov/abs/1106.3563}{{\tt
  arXiv:1106.3563}}].

\bibitem{KPT}
V.~Kumar, D.~S. Park, and W.~Taylor, {\it {6D supergravity without tensor
  multiplets}},  {\em JHEP} {\bf 1104} (2011) 080,
  [\href{http://xxx.lanl.gov/abs/1011.0726}{{\tt arXiv:1011.0726}}].

\bibitem{Heckman:2013kza}
J.~J. Heckman, {\it {Statistical Inference and String Theory}},
  \href{http://xxx.lanl.gov/abs/1305.3621}{{\tt arXiv:1305.3621}}.

\bibitem{Balasubramanian:2014bfa}
V.~Balasubramanian, J.~J. Heckman, and A.~Maloney, {\it {Relative Entropy and
  Proximity of Quantum Field Theories}},
  \href{http://xxx.lanl.gov/abs/1410.6809}{{\tt arXiv:1410.6809}}.

\bibitem{Green:1987mn}
M.~B. Green, J.~Schwarz, and E.~Witten, {\it {SUPERSTRING THEORY. VOL. 2: LOOP
  AMPLITUDES, ANOMALIES AND PHENOMENOLOGY}}, .

\bibitem{Intriligator20}
K.~A. Intriligator, {\it {Anomaly matching and a Hopf-Wess-Zumino term in 6d,
  N=(2,0) field theories}},  {\em Nucl.Phys.} {\bf B581} (2000) 257--273,
  [\href{http://xxx.lanl.gov/abs/hep-th/0001205}{{\tt hep-th/0001205}}].

\bibitem{MP}
D.~R. Morrison and D.~S. Park, {\it {F-Theory and the Mordell-Weil Group of
  Elliptically-Fibered Calabi-Yau Threefolds}},  {\em JHEP} {\bf 1210} (2012)
  128, [\href{http://xxx.lanl.gov/abs/1208.2695}{{\tt arXiv:1208.2695}}].

\bibitem{Mayrhofer:2012zy}
C.~Mayrhofer, E.~Palti, and T.~Weigand, {\it {U(1) symmetries in F-theory GUTs
  with multiple sections}},  {\em JHEP} {\bf 1303} (2013) 098,
  [\href{http://xxx.lanl.gov/abs/1211.6742}{{\tt arXiv:1211.6742}}].

\bibitem{Braun:2013yti}
V.~Braun, T.~W. Grimm, and J.~Keitel, {\it {New Global F-theory GUTs with U(1)
  symmetries}},  {\em JHEP} {\bf 1309} (2013) 154,
  [\href{http://xxx.lanl.gov/abs/1302.1854}{{\tt arXiv:1302.1854}}].

\bibitem{Cvetic:2013nia}
M.~Cvetic, D.~Klevers, and H.~Piragua, {\it {F-Theory Compactifications with
  Multiple U(1)-Factors: Constructing Elliptic Fibrations with Rational
  Sections}},  {\em JHEP} {\bf 1306} (2013) 067,
  [\href{http://xxx.lanl.gov/abs/1303.6970}{{\tt arXiv:1303.6970}}].

\bibitem{Antoniadis:2014jma}
I.~Antoniadis and G.~Leontaris, {\it {F-GUTs with Mordell-Weil U(1)'s}},  {\em
  Phys.Lett.} {\bf B735} (2014) 226--230,
  [\href{http://xxx.lanl.gov/abs/1404.6720}{{\tt arXiv:1404.6720}}].

\bibitem{Esole:2014dea}
M.~Esole, M.~J. Kang, and S.-T. Yau, {\it {A New Model for Elliptic Fibrations
  with a Rank One Mordell-Weil Group: I. Singular Fibers and Semi-Stable
  Degenerations}},  \href{http://xxx.lanl.gov/abs/1410.0003}{{\tt
  arXiv:1410.0003}}.

\end{thebibliography}\endgroup

\end{document}